\pdfoutput=1
\documentclass[fleqn, usenatbib]{mnras}

\newif\iflocal
\localfalse

\usepackage[T1]{fontenc}
\usepackage{ae,aecompl}
\usepackage{graphicx}
\usepackage{xspace}
\usepackage{amsmath}
\usepackage{framed} 
\usepackage{txfonts}
\usepackage{epstopdf}
\usepackage{color}
\usepackage{rotating}
\usepackage{natbib}
\usepackage{ulem}
\usepackage{xspace}
\usepackage{sistyle}
\SIthousandsep{,}

\iflocal

\newcommand{\kpc}{\>{\rm kpc}}

\newcommand{\msun}{\>{M_{\odot}}}

\newcommand{\yr}{\>{\rm yr}}

\def\gcm3{\mathrm{g} / \mathrm{cm}^3}

\def\rtoc{R_{\rm 200c}}

\def\mstar{M_{\ast}}

\def\colossus{\textsc{Colossus}\xspace}

\def\gtsima{$\; \buildrel > \over \sim \;$}
\def\ltsima{$\; \buildrel < \over \sim \;$}
\def\prosima{$\; \buildrel \propto \over \sim \;$}
\def\gsim{\lower.7ex\hbox{\gtsima}}
\def\lsim{\lower.7ex\hbox{\ltsima}}
\def\simgt{\lower.7ex\hbox{\gtsima}}
\def\simlt{\lower.7ex\hbox{\ltsima}}
\def\simpr{\lower.7ex\hbox{\prosima}}

\def\msun{{\rm M}_{\odot}}

\def\hi{${\rm H}$\,{\sc i}\xspace}
\def\htwo{${\rm H}_2$\xspace}
\def\hiht{\hi/${\rm H}_2$\xspace}

\def\sigmah{\Sigma_{\rm H}}

\def\sigmaht{\Sigma_{\rm H_2}}
\def\sigmahi{\Sigma_{\rm HI}}

\def\rhoh{\rho_{\rm H}}

\def\omegaht{\Omega_{\rm H_2}}
\def\omegahi{\Omega_{\rm HI}}

\def\fneutral{f_{\rm HI+H_2}}

\def\mg{M_{\rm gas}}
\def\mh{M_{\rm H}}
\def\mn{M_{\rm HI+H_2}}
\def\mht{M_{\rm H_2}}
\def\mhi{M_{\rm HI}}
\def\mstar{M_{*}}

\def\rhi{R_{\rm HI}}

\def\modell{\citetalias{leroy_08}\xspace}
\def\modelgk{\citetalias{gnedin_11}\xspace}
\def\modelk{\citetalias{krumholz_13}\xspace}
\def\modelgd{\citetalias{gnedin_14}\xspace}
\def\models{\citetalias{sternberg_14}\xspace}

\def\rtoc{R_{\rm 200c}}

\usepackage{etoolbox}
\makeatletter
\patchcmd{\NAT@citex}
  {\@citea\NAT@hyper@{\NAT@nmfmt{\NAT@nm}\NAT@date}}
  {\@citea\NAT@nmfmt{\NAT@nm}\NAT@hyper@{\NAT@date}}
  {}
  {}
\patchcmd{\NAT@citex}
  {\@citea\NAT@hyper@{%
     \NAT@nmfmt{\NAT@nm}%
     \hyper@natlinkbreak{\NAT@aysep\NAT@spacechar}{\@citeb\@extra@b@citeb}%
     \NAT@date}}
  {\@citea\NAT@nmfmt{\NAT@nm}%
   \NAT@aysep\NAT@spacechar%
   \NAT@hyper@{\NAT@date}}
  {}
  {}
\patchcmd{\NAT@citex}
  {\@citea\NAT@hyper@{%
     \NAT@nmfmt{\NAT@nm}%
     \hyper@natlinkbreak{\NAT@spacechar\NAT@@open\if*#1*\else#1\NAT@spacechar\fi}%
       {\@citeb\@extra@b@citeb}%
     \NAT@date}}
  {\@citea\NAT@nmfmt{\NAT@nm}%
   \NAT@spacechar\NAT@@open\if*#1*\else#1\NAT@spacechar\fi%
   \NAT@hyper@{\NAT@date}}
  {}
  {}
\makeatother

\else

\def\msun{{\rm M}_{\odot}}

\def\hi{${\rm H}$\,{\sc i}\xspace}
\def\htwo{${\rm H}_2$\xspace}
\def\hiht{\hi/${\rm H}_2$\xspace}

\def\sigmah{\Sigma_{\rm H}}

\def\sigmaht{\Sigma_{\rm H_2}}
\def\sigmahi{\Sigma_{\rm HI}}

\def\rhoh{\rho_{\rm H}}

\def\omegaht{\Omega_{\rm H_2}}
\def\omegahi{\Omega_{\rm HI}}

\def\fneutral{f_{\rm HI+H_2}}

\def\mg{M_{\rm gas}}
\def\mh{M_{\rm H}}
\def\mn{M_{\rm HI+H_2}}
\def\mht{M_{\rm H_2}}
\def\mhi{M_{\rm HI}}
\def\mstar{M_{*}}

\def\rhi{R_{\rm HI}}

\def\modell{\citetalias{leroy_08}\xspace}
\def\modelgk{\citetalias{gnedin_11}\xspace}
\def\modelk{\citetalias{krumholz_13}\xspace}
\def\modelgd{\citetalias{gnedin_14}\xspace}
\def\models{\citetalias{sternberg_14}\xspace}

\def\rtoc{R_{\rm 200c}}

\fi

\iflocal
\def\figdir{figs}
\else
\def\figdir{.}
\fi

\defcitealias{leroy_08}{L08}
\defcitealias{gnedin_11}{GK11}
\defcitealias{krumholz_09_kmt1}{KMT}
\defcitealias{krumholz_13}{K13}
\defcitealias{gnedin_14}{GD14}
\defcitealias{sternberg_14}{S14}

\defcitealias{diemer_18_hih2}{Paper I}
\defcitealias{calette_18}{C18}

\title[Atomic and molecular gas in IllustrisTNG]{Atomic and molecular gas in IllustrisTNG galaxies at low redshift}

\author[Diemer et al.]{Benedikt Diemer,$^{1}$\thanks{NHFP Einstein fellow; email: \href{mailto:benedikt.diemer@cfa.harvard.edu}{benedikt.diemer@cfa.harvard.edu}}
Adam R. H. Stevens,$^{2,3}$
Claudia del P. Lagos,$^{2,3,4}$
A. R. Calette,$^{5}$
\newauthor
Sandro Tacchella,$^{1}$
Lars Hernquist,$^{1}$
Federico Marinacci,$^{1,6}$
Dylan Nelson,$^{7}$
\newauthor
Annalisa Pillepich,$^{8}$ 
Vicente Rodriguez-Gomez,$^{9}$
Francisco Villaescusa-Navarro,$^{10}$
\newauthor
Mark Vogelsberger$^{11}$ \vspace{1mm}
\\
$^{1}$ Institute for Theory and Computation, Harvard-Smithsonian Center for Astrophysics, 60 Garden St., Cambridge, MA 02138, USA;\\
$^{2}$ International Centre for Radio Astronomy Research, The University of Western Australia, 35 Stirling Highway, Crawley, WA 6009, Australia\\
$^{3}$ ARC Centre of Excellence for All Sky Astrophysics in 3 Dimensions (ASTRO 3D)\\
$^{4}$ Cosmic Dawn Center (DAWN), Niels Bohr Institute, University of Copenhagen, N{\o}rregade, Copenhagen, Denmark DK-0000-0003-3631-7176 \\ 
$^{5}$ Instituto de Astronom\'ia, Universidad Nacional Aut\'onoma de M\'exico, A. P. 70-264, 04510, Ciudad de M\'exico, M\'exico \\
$^{6}$ Department of Physics and Astronomy, University of Bologna, via Gobetti 93/2, 40129 Bologna, Italy\\
$^{7}$ Max-Planck-Institut f{\"u}r Astrophysik, Karl-Schwarzschild-Str 1, D-85741 Garching, Germany\\
$^{8}$ Max-Planck-Institut f{\"u}r Astronomie, K{\"o}nigstuhl 17, D-69117 Heidelberg, Germany\\
$^{9}$ Instituto de Radioastronom\'{i}a y Astrof\'{i}sica, Universidad Nacional Aut\'{o}noma de M\'{e}xico, Apdo. Postal 72-3, 58089 Morelia, Mexico\\
$^{10}$ Center for Computational Astrophysics, Flatiron Institute, 162 5th Avenue, 10010, New York, NY, USA\\
$^{11}$ Department of Physics, Massachusetts Institute of Technology, Cambridge, MA 02139, USA
}

\date{Accepted 2019 May 8. Received 2019 April 22; in original form 2019 February 25}

\pubyear{2019}

\begin{document}
\label{firstpage}
\pagerange{\pageref{firstpage}--\pageref{lastpage}}
\maketitle


\begin{abstract}
We have recently developed a post-processing framework to estimate the abundance of atomic and molecular hydrogen (\hi and \htwo, respectively) in galaxies in large-volume cosmological simulations. Here we compare the \hi and \htwo content of IllustrisTNG galaxies to observations. We mostly restrict this comparison to $z \approx 0$ and consider six observational metrics: the overall abundance of \hi and \htwo, their mass functions, gas fractions as a function of stellar mass, the correlation between \htwo and star formation rate, the spatial distribution of gas, and the correlation between gas content and morphology. We find generally good agreement between simulations and observations, particularly for the gas fractions and the \hi mass-size relation. The \htwo mass correlates with star formation rate as expected, revealing an almost constant depletion time that evolves up to $z = 2$ as observed. However, we also discover a number of tensions with varying degrees of significance, including an overestimate of the total neutral gas abundance at $z = 0$ by about a factor of two and a possible excess of satellites with no or very little neutral gas. These conclusions are robust to the modelling of the \hiht transition. In terms of their neutral gas properties, the IllustrisTNG simulations represent an enormous improvement over the original Illustris run. All data used in this paper are publicly available as part of the IllustrisTNG data release.
\end{abstract}



\section{Introduction}
\label{sec:intro}

Over the last decade, large-volume cosmological simulations have matured into one of the key tools in the study of galaxy formation \citep[][see also the review of \citealt{somerville_15}]{schaye_10, schaye_15, dubois_14, vogelsberger_14_illustris, vogelsberger_14_nature, dave_16, dave_19}. However, such simulations achieve their large volumes at the cost of relatively low resolution, meaning that many physical processes need to be implemented through sub-grid prescriptions, which are inevitably based on a significant number of free parameters \citep[e.g.,][]{schaye_15, pillepich_18_tng}. These parameters must be calibrated to reproduce the properties of galaxies in the real Universe, such as the stellar mass function at low redshift. Galaxy properties that are not used in the tuning are then the simulations' predictions, and need to be carefully compared to observations to check whether the physical models create realistic galaxy populations. 

In this work, we focus on the IllustrisTNG simulation suite \citep{marinacci_18, naiman_18, nelson_18_color, pillepich_18, springel_18}, though we also briefly compare to the original Illustris-1 simulation \citep{vogelsberger_14_illustris, vogelsberger_14_nature, genel_14}. Many comparisons between the Illustris or IllustrisTNG simulations and observations have already been undertaken, including topics as wide-ranging as galaxy colours \citep{sales_15, nelson_18_color}, sizes \citep{genel_18}, the mass-metallicity relation \citep{torrey_18, torrey_19}, morphology \citep{snyder_15, rodriguezgomez_18, tacchella_19}, star formation \citep{sparre_15, diemer_17_sfh, donnari_19}, merger rates \citep{rodriguezgomez_15}, black holes \citep{weinberger_18, habouzit_19}, and the abundance of shell galaxies \citep{pop_18}.

However, one category of observations is noticeably under-represented in this list: the abundance, phase, and structure of gas in galaxies. For the original Illustris model, \citet{vogelsberger_14_nature} showed that the neutral fraction was broadly compatible with the trends seen in the Arecibo Legacy Fast ALFA (ALFALFA) survey. For the IllustrisTNG model, the total gas fraction in large groups and clusters was used as a constraint \citep[][see also \citealt{vogelsberger_18} and \citealt{barnes_18}]{pillepich_18_tng}. In addition, there have been studies of the larger-scale gas distribution around galaxies, namely of absorption lines due to the circum-galactic medium \citep{suresh_17, nelson_18_cgm}, the abundance of damped Lyman-alpha systems (DLAs) and the Lyman-alpha forest \citep{bird_14, gurvich_17}, hot gaseous haloes \citep{bogdan_15}, and other gas-related phenomena such as jellyfish galaxies \citep{yun_19_jellyfish}.

Nevertheless, we are still missing a complete census of the bulk of the observable gas in IllustrisTNG galaxies, namely, of atomic and molecular hydrogen (denoted \hi and \htwo hereafter). Especially the molecular (or simply, cold) component is of interest because it is thought to provide the fuel for star formation. While the IllustrisTNG simulations reproduce numerous properties of the observed stellar population, star formation is governed by an equilibrium interstellar-medium (ISM) model that does not directly predict the abundance of \htwo \citep{springel_03}. Thus, a detailed comparison of the gas and stellar properties of galaxies can tell us whether the relationship between gas and star formation is physically correct, or whether it emerges due to the tuning of the ISM model. There are, however, significant obstacles to \hi and \htwo comparisons on both the observational and theoretical sides.

Observationally, \hi in galaxies is detected either by the 21-cm emission from its spin-flip transition or by absorption in quasar or radio galaxy spectra \citep{zwaan_05, prochaska_05, allison_19}. The former technique has provided tight constraints on the local \hi abundance and its distribution in thousands of nearby galaxies \citep{martin_10, haynes_18}. At higher redshift, quasar sightlines provide rough constraints on the overall abundance of \hi, but we are lacking detailed information about \hi in galaxies \citep[e.g.,][]{zafar_13}. \htwo is, arguably, even more difficult to observe because the molecule lacks a permanent dipole moment \citep{draine_11}. Instead of observing \htwo directly, we rely on tracers such as CO, which are subject to uncertain conversion factors \citep{bolatto_13}. Nevertheless, molecular observations are now being pushed to high redshifts and high resolution by instruments such as ALMA, offering a chance to study gas in high-redshift galaxies in great detail \citep[e.g.,][]{walter_16}.

From a theoretical perspective, \hi and \htwo comparisons are challenging because the IllustrisTNG simulations predict the fraction of all gas that is in neutral hydrogen, but not whether this gas is atomic or molecular. The \hiht transition is governed by the balance between \htwo formation on dust grains and destruction by Lyman-Werner band ultraviolet (UV) photons, leading to complex dependencies on the density structure, star formation rate (SFR), and metallicity of the ISM \citep[e.g.,][]{spitzer_74, black_76, sternberg_89, elmegreen_89, elmegreen_93, krumholz_09_kmt1, draine_11}. These dependencies have been captured in a number of models, including observational correlations, calibrations from simulations, and analytic models \citep[e.g.,][]{blitz_06, gnedin_11, krumholz_13}. Such formulae can then be used to roughly constrain the molecular fraction in large-volume cosmological simulations in post-processing \citep[e.g.,][]{duffy_12}. \citet{lagos_15} applied this technique to the EAGLE simulation and showed that it reproduces the observed overall \htwo abundance, mass function, gas fraction, and depletion time. Based on similar modelling, \citet{bahe_16} investigated the abundance, size, and structure of \hi discs in EAGLE and found them to be broadly compatible with observations, though their conclusions depended somewhat on the chosen \hiht model \citep[for related results, see][]{marasco_16, lagos_16, crain_17, marinacci_17}.

In \citet[][hereafter \citetalias{diemer_18_hih2}]{diemer_18_hih2}, we undertook a systematic study of models that predict the \hi and \htwo fractions and applied them to the IllustrisTNG simulations. We extended the cell-by-cell modelling of \citet{lagos_15} by introducing a projection-based method and found the predictions of most models to be broadly compatible, as long as care is taken when computing the intermediate quantities they rely on. \citet{stevens_19_hi} used very similar modelling to investigate the dependence of \hi on environment, carefully mock-observing the simulation to take observational systematics into account. \citet{popping_19} use a version of the \hiht modelling that, among other differences, assumes a simpler approach to the UV field, and compare the high-redshift abundance of \htwo to ALMA observations \citep[see also][]{popping_14}.

In this work, we compare the results of \citetalias{diemer_18_hih2} to observations of galaxies at low redshift. We restrict ourselves to six observational metrics: the overall abundance of \hi and \htwo, their mass functions, gas fractions as a function of stellar mass, the correlation between \htwo and SFR, the spatial distribution of \hi and \htwo within galaxies, and correlations with morphology. Whenever possible, we carefully consider the observational systematics of the respective observational samples, attempting to faithfully mock-observe our simulated galaxies. All \hiht data used in this paper are publicly available as part of the IllustrisTNG data release \citep[][\href{http://www.tng-project.org/data/}{tng-project.org/data}]{nelson_19_datarelease}. 

The paper is structured as follows. We give an overview of all observational data used in Section~\ref{sec:obs}. We briefly review the simulation data in Section~\ref{sec:sim} (referring the reader to \citetalias{diemer_18_hih2} for details) and discuss our techniques for matching simulated galaxies to observations. We present the results of our comparisons in Section~\ref{sec:results}, including a brief review of the original Illustris-1 simulation. We further discuss particular successes and tensions in Section~\ref{sec:discussion} before summarizing our conclusions in Section~\ref{sec:conclusion}. We describe our experiments with morphology in Appendix~\ref{sec:app:morphology} and provide detailed tests of our methodology in Appendix~\ref{sec:app:apertures}.

We follow the notation of \citetalias{diemer_18_hih2}, denoting the masses of all gas as $\mg$, of all hydrogen as $\mh$, of atomic hydrogen as $\mhi$, of molecular hydrogen as $\mht$, and of neutral hydrogen as $\mn$. The same subscripts are used for volumetric mass densities (e.g., $\rhoh$), surface densities (e.g., $\sigmah$), number densities (e.g., $n_{\rm H}$), and column densities (e.g., $N_{\rm H}$). The neutral fraction, $\fneutral$, refers to the fraction of all gas (including helium and metals) that is in neutral hydrogen. 


\section{Observational Data}
\label{sec:obs}

In this section, we give an overview of the observational data used in our comparison. Each subsection discusses one of our six observational metrics.

\subsection{Overall Abundance of \hi and \htwo}
\label{sec:obs:omega}

As a first-order check on the abundance of neutral gas in IllustrisTNG, we use observations of the total \hi abundance as a function of redshift (though, for the remainder of the paper, we will mostly be concerned with $z \approx 0$). Our data compilation is largely based on that of \citet{rhee_18}, with a few sources added and removed. In particular, we use data from the blind ALFALFA 21-cm survey and similar observations \citep{zwaan_05, martin_10, freudling_11, hoppmann_15, obuljen_18}, from 21-cm observations of low-redshift galaxy samples \citep{lah_07, braun_12, delhaize_13, rhee_13, rhee_16, rhee_18}, and from quasar absorption lines studies that constrain the number of DLAs and Lyman-limit systems (LLSs) at high redshift \citep{prochaska_05, rao_06, rao_17, noterdaeme_09, noterdaeme_12, zafar_13, neeleman_16, bird_17}.
 
The \htwo abundance of the Universe at low redshift can currently not be determined in blind surveys because the emission of the CO tracer is simply too weak. Similar caveats affect all \htwo-related comparisons in this paper: \htwo is never observed directly, and the relationship between \htwo and CO is probably much more complicated than suggested by constant conversion factors \citep{wolfire_10, glover_11, leroy_11, narayanan_11, narayanan_12, feldmann_12, bolatto_13}. Thus, all \htwo abundances are to be interpreted as estimates that are likely accurate to factors of a few. 

Subject to this caveat, we compare our simulations to low-redshift constraints based on targeted CO observations of galaxies \citep{keres_03, obreschkow_09a, saintonge_17}. While the first blind surveys are beginning to probe the abundance of \htwo at high redshift \citep{walter_16, decarli_16_aspecs1, pavesi_18, riechers_19}, we do not use their results in this study because the small volumes probed lead to significant sample variance and other complications (see \citet{popping_19} for a detailed comparison). In recent years, the \htwo abundance is increasingly being inferred from the dust continuum emission, e.g., as measured by ALMA, leading to different systematics \citep{scoville_17, tacconi_18}. However, such measurements are currently possible only at high redshift, and are thus not the focus of this paper.

\subsection{Mass Functions}
\label{sec:obs:mfunc}

The \hi mass function has been measured to exquisite accuracy by ALFALFA \citep{giovanelli_05, martin_10, haynes_11, huang_12, jones_18, haynes_18}. The final $\alpha$.100 catalogue contains about \num{30000} sources below $z = 0.06$. Our primary constraint for the \htwo mass function is the work of \citet{boselli_14_hrs2}, specifically their estimate based on a fixed CO-to-\htwo conversion factor (which we convert to $\alpha_{\rm CO} = 2.0 \msun / ({\rm K\ km/s\ pc}^2)$ as in \citealt{popping_19}). The underlying data are taken from the {\it Herschel} Reference Survey, a sample of 323 galaxies observed with the SPIRE instrument \citep{boselli_10}. The galaxies in this survey are local, with distances between 15 and 25 Mpc. For comparison, we also show the \htwo mass function of \citet{obreschkow_09a}, which is based on \citet{keres_03}. The underlying dataset is the FCRAO survey \citep{young_95}, which measured the CO emission for $300$ local galaxies. However, due to the range of distances and observing strategies, this survey is somewhat difficult to mock-observe.

\subsection{Gas Fractions}
\label{sec:obs:fraction}

By gas fractions, we mean the ratio of the mass in \hi or \htwo divided by stellar mass. These ratios are commonly constrained observationally and help us determine whether IllustrisTNG produces an appropriate amount of gas in galaxies of a certain stellar mass. The scatter in the gas fractions is known to be sizeable, meaning that both the median fraction and the distribution around the median are of interest. However, comparing simulations to observed gas fractions is non-trivial because of numerous survey-specific aspects such as the galaxy selection, aperture (or beam), and other assumptions used in the processing of the observations. \citet{stevens_19_hi} recently demonstrated that the total \hi fractions in IllustrisTNG can differ significantly from mock observations that take the observational systematics into account.

To avoid having to mock-observe many heterogeneous surveys, we rely on the recent compilation of \citet[][hereafter \citetalias{calette_18}]{calette_18}. They quantify the \hi and \htwo fractions as a function of stellar mass, including fitting functions for the mass-dependent distribution of gas fractions. \citetalias{calette_18} amalgamate \hi and \htwo masses from a large range of radio observations that have been combined with stellar luminosities or masses from optical or infrared observations. In particular, their \hi data includes the Updated Nearby Galaxy Catalogue \citep{karachentsev_13}, GASS \citep{catinella_13, catinella_18}, the {\it Herschel} Reference Survey \citep{boselli_10, boselli_14_hrs1, boselli_14_hrs2, boselli_14_hrs3}, ATLAS$^{\rm 3D}$ \citep{serra_12}, the Nearby Field Galaxy Survey \citep{jansen_00_a, jansen_00_b, kannappan_13}, THINGS \citep{leroy_08}, ALFALFA \citep{huang_12_dwarfs}, UNAM-KIAS \citep{hernandeztoledo_10}, AMIGA \citep{lisenfeld_11}, and a number of other compilations \citep{geha_06, stark_13, bradford_15}. Their \htwo data is taken from the {\it Herschel} Reference Survey, COLD GASS \citep{saintonge_11_coldgass}, ATLAS$^{\rm 3D}$, HERACLES \citep{leroy_08}, ALLSMOG \citep{bothwell_14}, EGNoG \citep{bauermeister_13}, and \citet{stark_13}. These data sources are discussed in detail in Appendices A and B of \citetalias{calette_18}.

Using the \citetalias{calette_18} compilation offers a number of advantages. First, \citetalias{calette_18} homogenize the survey data to a \citet{chabrier_03} stellar initial mass function (IMF), the same IMF used in IllustrisTNG. Second, numerous survey biases and upper limits are taken into account, the latter via a Kaplan-Meier estimator. For example, surveys can easily be biased towards gas-rich galaxies (see Fig.~1 in \citetalias{calette_18} for an example of a fitting function that overestimates the full, debiased data set). Third, all CO observations are converted to \htwo masses based on the same mass-dependent $x_{\rm CO}$ factor.

The \citetalias{calette_18} compilation does, however, introduce one complication: they split the galaxy sample into early and late-type galaxies (ETGs and LTGs, respectively). This distinction is sensible because the gas content of galaxies strongly correlates with their morphological type \citep{kannappan_13, boselli_14_hrs3}. Thus, quantifying the median gas fraction of all galaxies introduces additional scatter compared to the relations split by morphology. In Section~\ref{sec:sim:morphology} and Appendix~\ref{sec:app:morphology}, we experiment with different morphological selections, but find no criterion that reliably creates late-type and early-type samples with the properties expected from observations. To avoid the resulting difficulties in interpretation, we reconstruct the distribution of gas fractions in the overall galaxy sample. \citetalias{calette_18} provide fitting functions for the distribution of gas fractions of early and late types as a function of stellar mass (their equations 3 and 7), as well as the ETG fraction as a function of stellar mass. We add the distributions of the LTG and ETG samples accordingly. For the ETG sample, \citetalias{calette_18} give a mass-dependent cutoff gas fraction below which the distribution cannot be reliably inferred because the majority of observations are upper limits. We separately constrain the properties of those galaxies with gas fractions above the limit as well as the fraction of galaxies below the limit. We use a slightly updated version of the \citetalias{calette_18} best-fit parameters. 

\citetalias{calette_18} find that, once they split into LTGs and ETGs and homogenize the data, the different samples are broadly consistent, except for certain datasets where strong biases are expected. Those are collected in a ``bronze'' sample, which we do not use in our analysis. \citetalias{calette_18} show that their gas fractions are compatible with the observed mass functions and that the \hi fractions measured in individual LTGs are compatible with the stacked ALFALFA analysis of \citet{brown_15}, indicating that the galaxy sample is not missing any major contributions to the overall \hi abundance. Given that some data sources were selected from optical samples such as SDSS, there may be a bias against satellites close to their central. For example, galaxies in the SDSS-I and II spectroscopic surveys could not be closer than 55'' due to fibre collisions \citep[e.g.,][]{zehavi_02, guo_12}. We will return to this difficulty when discussing the gas fractions of satellites in Sections~\ref{sec:results:fraction} and \ref{sec:discussion:gasfree}. \citetalias{calette_18} assume $h = 0.7$ while IllustrisTNG uses $h = 0.6774$; we rescale masses wherever appropriate.

\subsection{Correlations with Star Formation Rate}
\label{sec:obs:sfr}

For our comparison of \htwo mass and SFRs, we use the xCOLD GASS survey \citep{saintonge_11_coldgass, saintonge_17}. This dataset is composed of two surveys, COLD GASS and xCOLD GASS-low, which combine to cover all stellar masses above $10^9 \msun$. The CO observations were taken with the IRAM telescope, their stellar properties are taken from SDSS and other surveys. It is well established that the depletion time, $t_{\rm depl} \equiv \mht / {\rm SFR}$, is roughly constant but depends on redshift \citep{leroy_08, leroy_13, bigiel_08, bigiel_11, tacconi_10, tacconi_13, tacconi_18, daddi_10_laws, genzel_10, genzel_15, saintonge_11_depletion, saintonge_11_coldgass, saintonge_17}. We compare IllustrisTNG to the parameterization of \citet{tacconi_18}, $t_{\rm depl} \approx 1\ {\rm Gyr} \times (1 + z)^{-0.57}$.

\subsection{Spatial Distribution of Gas}
\label{sec:obs:sizes}

One of the most well-constrained indicators of the spatial distribution of gas in galaxies is the \hi radius, $\rhi$, commonly defined as the radius where the surface density falls below $\sigmahi < 1 \msun/{\rm pc}^2$. This radius forms a tight power-law relationship with $\mhi$, at least in the local galaxy samples where it has been observed in resolved 21-cm observations \citep{broeils_94, broeils_97, verheijen_01_masssize, wang_13_bluedisks, wang_14_hi, martinsson_16, ponomareva_16, wang_16_hi}. We use the data compilation of \citet{wang_16_hi}, who combine $\rhi$--$\mhi$ data from about 500 galaxies measured by 15 different observational projects. For comparison, we also show the similar relation of \citet{lelli_16}. The measured \hi sizes correspond to a face-on orientation.

The resolved \hi observations used here do not represent a volume-limited sample but are typically selected to include mostly \hi-rich targets. However, all morphological types are present in the surveys and the gas fractions roughly match the averages observed in volume-limited samples, indicating that there are no strong biases in the selection \citep{lelli_16}. Moreover, the $\rhi$--$\mhi$ is extremely robust in that it does not vary as a function of stellar brightness or stellar diameter \citep{wang_16_hi, lelli_16}.

Going beyond $\rhi$, we also investigate the radial dependence of the \hi surface density. For this purpose, we use the \hi profiles from the Bluedisk survey \citep{wang_13_bluedisks, wang_14_hi}. They initially selected 25 galaxies as a control sample and 25 galaxies as a \hi-rich sample, but the surface density profiles (with radii scaled to $\rhi$) turn out to be almost identical. The Bluedisk galaxies have $10^{10} \msun < M_* < 10^{11} \msun$ and lie on the star-forming main sequence, with SFRs between $0.5$ and $10\ \msun/{\rm yr}$ \citep{cormier_16}.
 
Finally, we wish to quantify the spatial distribution of \htwo, but there are a number of issues. First, the observational sample sizes are limited by the difficulty of obtaining spatially resolved CO observations. Second, the radial profiles of \htwo appear to be diverse \citep[e.g.,][]{young_82, regan_01, leroy_08, leroy_09}. Third, uncertain (and perhaps radially varying) CO-to-\htwo conversion factors complicate the interpretation of observations \citep[][and references therein]{bolatto_13}. Given these difficulties, we delegate a detailed comparison with CO profiles and maps to future work. Instead, we restrict ourselves to a comparison of the observed half-mass radii of CO and stars from the EDGE-CALIFA survey \citep{bolatto_17}. This dataset adds spatially resolved CARMA CO observations to a subset of the optical integral field unit spectroscopy from CALIFA \citep{sanchez_12, walcher_14}. We note that there are numerous other ways to quantify the CO spatial scale, but some of them use relatively high surface densities that are difficult to reproduce in our simulations. For example, \citet{davis_13_extent} measured $R_{\rm CO}$ as the radius where the surface density of \htwo reaches $15 \, \msun / {\rm pc}^2$, corresponding to their $3 \sigma$ detection limit. This surface density is never reached by two-thirds of TNG100 galaxies with $\mstar > 10^{10} \msun$. As a result, \citet{davis_13_extent} find very small values for $R_{\rm CO}$ that are close to the force resolution scale used in IllustrisTNG.


\section{Simulation Data}
\label{sec:sim}

Our modelling of the \hiht transition and the relevant details of the IllustrisTNG simulations were discussed at length in \citetalias{diemer_18_hih2}. Here we briefly review the aspects that are most pertinent to this work and refer to the reader to \citetalias{diemer_18_hih2} for details.

\subsection{The IllustrisTNG Simulations}
\label{sec:sim:illustris}

The IllustrisTNG suite of cosmological magneto-hydrodynamical simulations was run using the moving-mesh code \textsc{Arepo} \citep{springel_10}. In this paper, we use the highest-resolution versions of the $100$ and $300$ Mpc box sizes, hereafter referred to as TNG100 and TNG300 \citep{marinacci_18, naiman_18, nelson_18_color, pillepich_18, springel_18}. As the TNG100 simulation has significantly higher spatial and mass resolution than TNG300, we interpret differences in their galaxy properties to be resolution effects, i.e., we trust the TNG100 results over TNG300 (see also \citetalias{diemer_18_hih2}). We use a lower-resolution version of TNG100, TNG100-2, to quantify the convergence with resolution.

IllustrisTNG adopts the \citet{planck_16} cosmology: $\Omega_{\rm m} = 0.3089$, $\Omega_{\rm b} = 0.0486$, $h = 0.6774$, and $\sigma_8 = 0.8159$; the same cosmology is assumed throughout the paper. The simulations follow a large range of physical processes, including prescriptions for gas cooling, star formation, stellar winds, metal enrichment, supernovae, black hole growth, and active galactic nuclei \citep{weinberger_17, pillepich_18_tng}. This so-called TNG model builds on the original Illustris model \citep{vogelsberger_13, vogelsberger_14_illustris, vogelsberger_14_nature, genel_14, torrey_14, sijacki_15}. We briefly compare our results to the original simulation in Section~\ref{sec:results:orig}. Dark matter haloes and their galaxies were identified using the \textsc{Subfind} algorithm \citep{springel_01_subfind}.

\subsection{Modelling the {\rm \hiht} Fractions}
\label{sec:sim:hih2}

In \citetalias{diemer_18_hih2}, we considered the \hiht models of \citet{leroy_08}, \citet{gnedin_11}, \citet{krumholz_13}, \citet{gnedin_14}, and \citet{sternberg_14} (hereafter \modell, \modelgk, \modelk, \modelgd, and \models, respectively). These models represent three separate categories: (i) observational correlations with the mid-plane pressure of galaxies (\modell), (ii) fitting functions trained on high-resolution simulations with radiative transfer and chemical networks (\modelgk, \modelgd), and (iii) analytical models of simplified, individual clouds that were modified to represent an average over a larger ISM region (\modelk, \models). Models in the latter two categories demand the Lyman-Werner band UV field strength as an input parameter, and we thus modelled the UV field by propagating light from young stars in an optically thin fashion. 

Conventionally, \hiht models have been computed cell by cell, i.e., by assigning a molecular fraction to each gas element. The main issue with this method is that all models listed above rely on surface densities rather than volume densities. For a 3D gas cell, such surface densities are commonly estimated by multiplying with the Jeans length, roughly representing the size of a self-gravitating system in equilibrium \citep[e.g.,][]{schaye_01}. In \citetalias{diemer_18_hih2}, we showed that this approximation does not, in general, reproduce the actual surface density of our galaxies. We proposed an alternative way to compute the models in projection, where each galaxy is rotated into a face-on position, projected onto a 2D map, and the model equations are solved in two dimensions. 

In this work, we remain agnostic as to which method and which \hiht model is most accurate and view their spread as a systematic uncertainty. We consider both the cell-by-cell and projected versions, with the exception of the cell-by-cell \modell model, which was shown to be unphysical \citepalias{diemer_18_hih2}. Thus, the \hi and \htwo masses of each galaxy are estimated by a total of nine different methods. In the forthcoming figures, we shade the region between the models but do not show the individual models to avoid crowding. The differences between the model predictions are shown in the corresponding figures in \citetalias{diemer_18_hih2}.

\subsection{Galaxy Selection}
\label{sec:sim:selection}

In \citetalias{diemer_18_hih2}, we computed the \hiht fractions of all TNG100 galaxies with either $M_* > 2 \times 10^8 \msun$ or $M_{\rm gas} > 2 \times 10^8 \msun$. We use the same mass limits in this work, but increase them to $M_* > 5 \times 10^{10} \msun$ and $M_{\rm gas} > 5 \times 10^9 \msun$ for TNG300. These selections result in stellar-mass-selected samples of \num{43213} and \num{22109} galaxies for TNG100 and TNG300, respectively, and \num{140923} and \num{449176} galaxies for the gas-mass-selected samples. We caution that our mass cut-offs in TNG100 are chosen somewhat aggressively, as they corresponds to about $200$ stellar-population particles or gas cells. 

We do not cut the galaxy sample any further at this point, though we will later impose additional cuts to mimic the galaxy selection of various observational datasets. We include both central and satellite galaxies because they are not generally separated in observations. As a result, our sample contains galaxies from different environments such as field galaxies and cluster members. It is now well established that the \hi fractions tend to be significantly lower in the latter, a trend partially captured in the morphology-density relation where ETGs tend to both live in denser environments and contain less gas \citep{haynes_84, cortese_11, catinella_13, boselli_14_hrs3, stevens_19_hi}. Similarly, different gas fractions have been observed in the satellite population by splitting observed samples using group catalogues \citep{brown_17}. However, \citetalias{calette_18} do not split their galaxy sample by any isolation criterion, and the mass functions also include both satellites and centrals. Moreover, the IllustrisTNG simulations have been shown to recover the various trends with environment \citep{stevens_19_hi}.

Finally, we do not cut out objects with unusual ratios of their gas, stellar, and dark matter content. For example, IllustrisTNG contains some objects with barely any dark halo which can be traced to non-cosmological formation mechanisms \citep{nelson_19_datarelease}. We find that such objects contribute to a population of gas-free galaxies that we discuss further in Sections~\ref{sec:results:fraction} and \ref{sec:discussion:gasfree}.

\subsection{Mock-observing Masses, Sizes, and SFRs}
\label{sec:sim:apertures}

The most natural definition for the masses of stars or gas in simulated galaxies is to include all particles or gas cells that are gravitationally bound to the respective halo or subhalo. When comparing to observations, however, this definition rarely applies because the observed data are usually based on a finite beam size or aperture, which can either reduce the measured gas mass due to partial coverage or increase it due to confusion with other objects. Such effects can lead to significant differences in the inferred gas masses \citep[e.g.,][]{stevens_19_hi}.

We compute all aperture masses from linearly interpolated, cumulative mass profiles of the respective species. The profiles contain only matter bound to the respective subhalo and are binned in $50$ linearly spaced radii at fixed fractions of the outer radius of our 2D maps (equation~4 in \citetalias{diemer_18_hih2}). If the desired aperture is larger than the largest radial bin, we use the cumulative mass in that final bin. For any quantity that is defined cell-by-cell or particle-by-particle (e.g., stellar mass or gas masses), we could use a spherically averaged (3D) or a projected (2D) profile. For disc-like, anisotropic systems, this choice could make a difference in the computed masses. We choose to compute all masses and sizes from projected profiles to be consistent with those quantities that are computed in projection in the first place (Section~\ref{sec:sim:hih2}). In such cases, we cannot use 3D profiles because the information about the 3D positions has been lost. We have tested the difference between using 2D and 3D profiles and find it to be insignificant, with no clearly discernible trend in the inferred masses.

We compute sizes in a similar way, namely, we linearly interpolate the projected mass profile to find a density threshold (e.g., $\sigmahi = 1 \msun / {\rm pc}^2$) or a fraction of the integrated mass (e.g, the half-mass radius). This method is compatible with the observational samples to which we compare, where sizes correspond to a face-on orientation (Section~\ref{sec:obs:sizes}).

\subsubsection{Stellar Masses and Sizes}
\label{sec:sim:apertures:star}

Optical surveys used to measure stellar masses tend to underestimate the true extent of the stellar distribution \citep[e.g.,][]{bernardi_13, kravtsov_14}. To mimic such effects, we define $M_*$ as the stellar mass within a fixed aperture of $30 \kpc$. \citet{schaye_15} demonstrated that this definition tracks the Petrosian radii used in observations \citep[see also][for an investigation of various definitions in IllustrisTNG]{pillepich_18}. A fixed $30$ kpc aperture also matches the BaryMP technique of \citet{stevens_14} to about 20\% accuracy (with larger deviations at $M_* \gtrsim 10^{11} \msun$). 

In extremely rare cases ($0.003$\% of galaxies in TNG100, $0.02$\% in TNG300), the radius of the innermost bin of the stellar mass profile is larger than $30 \kpc$. In those cases, we linearly interpolate the logarithmic value of the first bin from zero. We find that the details of this procedure have no appreciable impact on our results. We have also tested that increasing the number of bins by more than two-fold has less than a one-percent effect on the computed stellar masses.

Besides the bias related to the size of a galaxy, stellar masses suffer from significant systematic uncertainties such as the assumed IMF and the modelling of the stellar light profiles. For example, while \citetalias{calette_18} homogenize their data to the same IMF, modelling uncertainties largely remain (see their Section~2.1.3). Thus, we add a log-normal scatter of $0.2$ dex to the stellar masses from IllustrisTNG \citep{marchesini_09, mancini_11, bower_12, mitchell_13}. When comparing gas fractions at fixed stellar mass, this has two main effects. First, features in the relations are smoothed out, which can hide trends such as rapid changes due to the feedback implementation (see the discussion in \citealt{stevens_19_hi} and \citealt{nelson_19_datarelease}). Second, as the gas fractions typically fall with stellar mass, Eddington bias can shift the gas fractions upwards \citep{eddington_13}. 

\subsubsection{Star Formation Rates}
\label{sec:sim:apertures:sfr}

We compare our \htwo masses and SFRs to the xCOLD GASS survey, which combines measurements of \hi and CO properties of galaxies with optical surveys \citep{saintonge_17}. Their SFRs are derived using a combination of UV and infrared (IR) luminosities \citep{janowiecki_17}. While UV tracers such as the H$\alpha$ line are indicative of star formation on short timescales, IR tracers are sensitive to timescales of a few hundred Myr; their combination typically traces variations over roughly $50$--$150$ Myr \citep[e.g.,][]{caplar_19}.

We use the instantaneous SFRs given by the simulation, but caution that there may be a slight systematic offset \citep{donnari_19}. To put a simplistic upper limit on this offset, we have compared the instantaneous SFRs of IllustrisTNG galaxies, defined as the sum of the SFRs of a galaxy's gas cells, to an average over the stars formed within about 200 Myr. Given the timescales discussed above, this represents a conservative estimate of any differences. For the relevant stellar mass and redshifts, $z = 0$ and $z = 2$, we find that the median instantaneous SFRs of TNG100 galaxies differ from the averaged ones by about 3\%, a negligible offset. There is, however, a scatter of $0.15$ dex at $z = 0$ and $0.6$ dex at $z = 2$ to which we return when interpreting our results in Section~\ref{sec:results:sfr}.

\subsubsection{\hi Masses and Sizes}
\label{sec:sim:apertures:hi}

The majority of the \hi data described in Section~\ref{sec:obs} are based on Arecibo observations. The beam width of the Arecibo telescope at 21 cm is about $3.4$' \citep[e.g.,][]{jones_18}, corresponding to a physical aperture that depends on redshift. For example, at the median redshift of the GASS survey, $z = 0.037$, $3.4$' corresponds to roughly $140 \kpc$, leading \citet{bahe_16} to choose a fixed aperture radius of  $70 \kpc$. The median redshift of the ALFALFA sample is slightly lower, which would demand a slightly smaller aperture. Moreover, only \hi within a certain range of relative velocity would be registered in the 21-cm band. However, \citet{bahe_16} found that a spherical aperture of $70 \kpc$ corresponds very well to a more complex procedure that takes the velocity of the gas into account (see their Appendix~A). We thus follow their suggestion and use a circular $70 \kpc$ aperture in face-on projection for all our \hi masses. As Arecibo is a single-dish telescope, we apply a Gaussian beam instead of a radial cut-off, i.e., we weight the projected \hi density profile by a Gaussian with $\sigma = 70$ kpc. The differences are negligible at all stellar masses except at $M_* > 10^{11}\msun$ where using the Gaussian slightly increases the \hi mass. We further discuss the impact of the chosen aperture in Appendix~\ref{sec:app:apertures}. 

One observational complication that is not taken into account is blending: galaxies sometimes lie behind each other and close enough in velocity space to be confused into a single detection, increasing the measured \hi fraction and decreasing the number of objects. However, \citet{jones_15} investigated this effect for ALFALFA and found that it would change the ALFALFA mass function by less than $3 \sigma$, a small shift given the extremely tight error bars. In Appendix~\ref{sec:app:apertures}, we compare our \hi masses to those of \citet{stevens_19_hi} who take blending into account and find that the differences do not appreciably change our results. For our \htwo masses, blending effects are negligible due to the small spatial extent of \htwo and due to the optical galaxy selection (Sections~\ref{sec:obs:fraction} and \ref{sec:sim:apertures:h2}).

Finally, in Section~\ref{sec:results:size}, we use the \hi radius $\rhi$, defined as the radius where the surface density of \hi falls below $1 \msun / {\rm pc}^2$. We measure this radius from projected one-dimensional density profiles because the observational measurements are face-on equivalents, i.e., they are corrected for inclination. As discussed in Section~\ref{sec:obs:sizes}, the datasets are heterogeneous but are generally very well resolved and include the entire \hi disc. Thus, we do not impose an aperture when measuring $\mhi$ for use in the $\rhi$--$\mhi$ relation.

\subsubsection{\htwo Masses and Sizes}
\label{sec:sim:apertures:h2}

For \htwo, the chosen aperture turns out to be much more important than for \hi. The data used in \citetalias{calette_18} are dominated by the xCOLD GASS survey, which consists of $532$ measurements of the CO luminosity taken with the IRAM 30 m telescope \citep{saintonge_17}. The 3 mm channel (which contains the CO 1--0 transition) has a field of view of 22''. Over the survey's redshift range, $0.01 < z < 0.05$, this corresponds to between $4.6$ and $20$ kpc. Based on the high-mass end of this redshift range, we use $20$ kpc as an approximation for the typical beam of IRAM. The instrument used is a single-pixel bolometer, meaning that the beam is Gaussian in shape. The quoted beam size corresponds to the full width at half maximum, or $2.355 \sigma$, meaning that the standard deviation of our Gaussian aperture is $\sigma \approx 8.5$ kpc. \citet{saintonge_17} apply a correction factor for inclination which increases the measured CO luminosity by a median factor of $1.17$ \citep[see also][]{saintonge_12}. Since we mock-observe galaxies in face-on orientation, this correction should roughly match our procedure.

The {\it Herschel} Reference Survey observations that underlie our \htwo mass function use a rather different instrument, SPIRE \citep{boselli_10}. The SPIRE field of view is 4' by 8', which we approximate by a single aperture of 6'. The survey sources lie between 15 and 25 Mpc; we take 20 Mpc as a typical distance. At this separation, 6' corresponds to $36$ kpc or a radial aperture of $18$ kpc. Rather than a single beam such as IRAM, SPIRE has a resolution of 30''. Thus, we use a hard cutoff radius of $18$ kpc rather than a Gaussian aperture, though the difference is insignificant.

Returning to the IRAM-based observations used in the gas fraction measurements, we might worry that the \htwo mass of our sample could be overestimated because a fraction of the galaxies lies at lower redshift and experiences even smaller apertures. However, there are also reasons to fear that our measurement might underestimate the observations because of the spatial extent of \htwo in IllustrisTNG. In Section~\ref{sec:results:size:h2}, we demonstrate that the \htwo in our simulated galaxies tends to be more spread out than suggested by CO observations, both in relation to the stellar component and in absolute units. This disagreement begs the question of whether we wish to mock-observe exactly what a telescope would see given the simulation data, or whether we are interested in quantifying the overall \htwo mass of the simulated galaxies. For the purposes of gas fractions and mass functions, we are interested in the latter. Thus, our results should probably be seen as a lower limit on the \htwo mass one would observe given the simulations. We quantitatively investigate the effect of the \htwo aperture in Appendix~\ref{sec:app:apertures}.

\subsection{Separation into Early and Late Types}
\label{sec:sim:morphology}

As discussed in Section~\ref{sec:obs:fraction}, we base our comparisons of gas fractions on the observational compilation of \citetalias{calette_18}, who split galaxies into LTGs and ETGs. This distinction is physically meaningful because the \hi and \htwo fractions of the two groups differ significantly, with ETGs containing less gas than LTGs at fixed stellar mass. To assess whether IllustrisTNG galaxies match this trend, we split our sample into LTGs and ETGs based on the concentration of stellar mass, $C_{80,20} > 4.9$. This split reproduces the \citetalias{calette_18} ETG fraction for both centrals and satellites. Here, we define concentration as $C_{80,20} \equiv 5 \log_{10} (r_{80} / r_{20})$, where $r_{80}$ and $r_{20}$ are the radii enclosing 80\% and 20\% of the gravitationally bound stellar mass, respectively. This parameter (observationally defined through stellar light rather than mass) is known to strongly correlate with the morphological type of galaxies \citep{kent_85, bershady_00, lotz_04}. The concentration in IllustrisTNG appear to match observations reasonably well \citep{rodriguezgomez_18}, especially when a scatter in stellar mass is applied. In Appendix~\ref{sec:app:morphology}, we describe our experiments with morphology and investigate the correlation between various morphological indicators and other galaxy properties.


\section{Results}
\label{sec:results}

\begin{figure}
\centering
\includegraphics[trim =  3mm 4mm 1mm 0mm, clip, scale=0.7]{\figdir/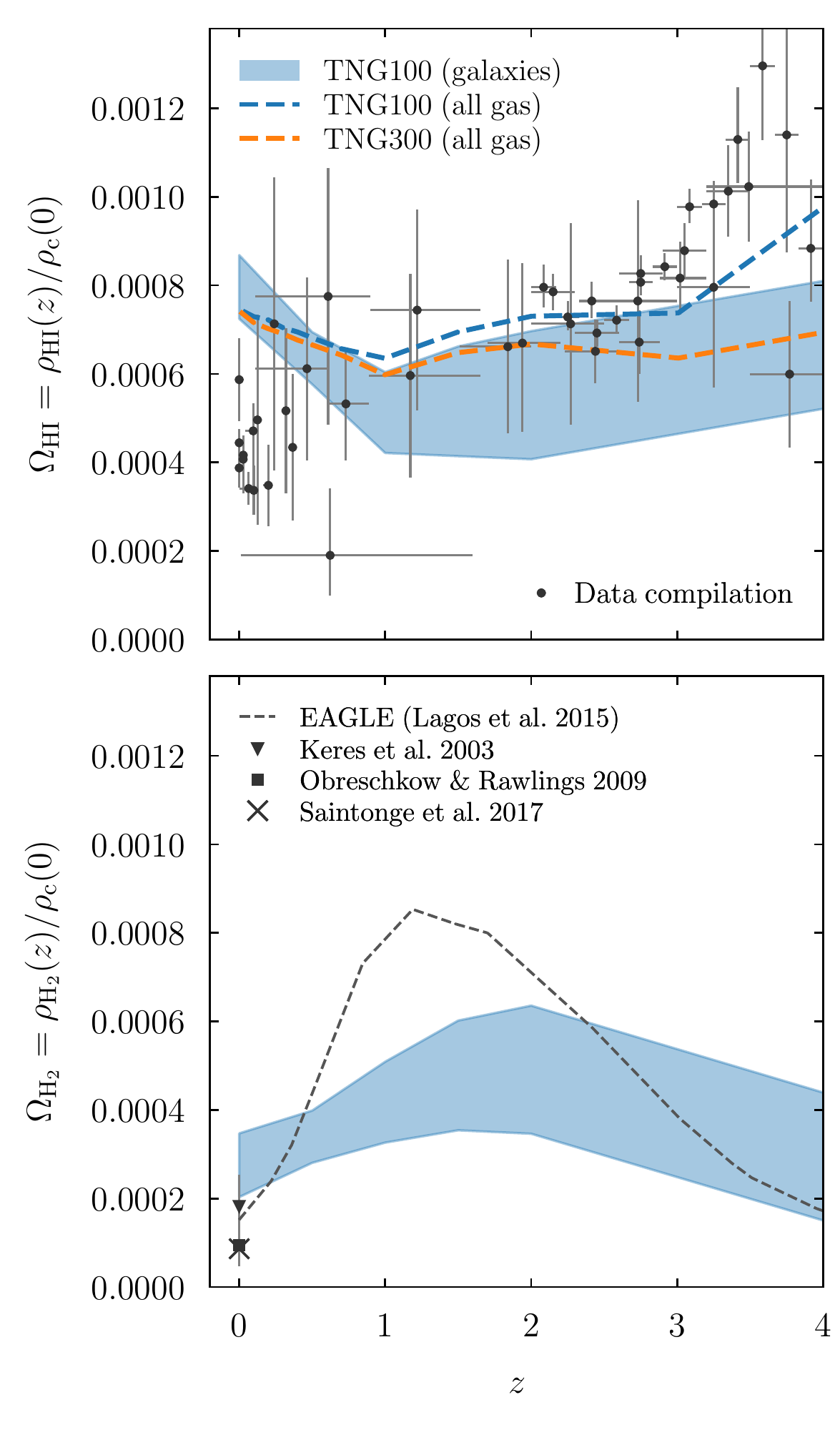}
\caption{Overall abundance of \hi (top) and \htwo (bottom) as a function of redshift. As in all following figures, the shaded blue region encloses the curves due to the different \hiht models. The TNG100 result represents a lower limit because it does not include galaxies outside of our selection (${\rm max}(M_{\rm gas}, M_*) \geq 2 \times 10^8 \msun$, Section~\ref{sec:sim:selection}) or gas that is not associated with any subhalo. However, the dashed lines in the top panel demonstrate that using all gas particles gives similar results \citep{villaescusanavarro_18}. We expect the amount of \htwo outside galaxies to be negligible. See Section~\ref{sec:obs:omega} for details on the \hi data compilation.}
\label{fig:omega}
\end{figure}

We are now ready to compare the gas content of IllustrisTNG galaxies to observations. We split this section into six observational metrics, namely the overall abundance of \hi and \htwo (Section~\ref{sec:results:omega}), mass functions (Section~\ref{sec:results:mfunc}), gas fractions (Section~\ref{sec:results:fraction}), correlations with SFR (Section~\ref{sec:results:sfr}), the spatial distribution of gas (Section~\ref{sec:results:size}), and the correlation between neutral gas content and morphology (Section~\ref{sec:results:morphology}). Throughout the paper, we adhere to a consistent plotting scheme where blue lines and shapes correspond to TNG100, orange to TNG300, and gray or black elements indicate observational data. Wherever we show \hiht-related quantities, we indicate the variation between the nine different models as a dark shaded area. We omit the individual model lines to avoid confusion and refer the interested reader to \citetalias{diemer_18_hih2} where the detailed differences between the models are shown. Where applicable, the 68\% scatter is shown as a lighter shaded area of the same colour. In this case, it would not make sense to show the maximum scatter of the models; instead, we take the mean of the 16th and 84th percentiles of all nine models. We focus on median rather than mean quantities to avoid issues with outliers and galaxies that contain no or very little gas. To avoid overcrowding the figures in this section, we do not show results from the original Illustris-1 simulation or from the lower-resolution TNG100-2 simulation. However, we provide full sets of figures comparing TNG100 to Illustris-1 and to TNG100-2 at \href{http://www.benediktdiemer.com/data/}{benediktdiemer.com/data}. We briefly review the results for Illustris-1 in Section~\ref{sec:results:orig}.

\subsection{Overall Abundance of \hi and \htwo}
\label{sec:results:omega}

To set a baseline expectation for the match between the neutral gas content in IllustrisTNG galaxies and the real Universe, we first consider the overall abundance of \hi and \htwo as a function of redshift as shown in Fig.~\ref{fig:omega}. These abundances, denoted $\omegahi$ and $\omegaht$, are defined with respect to the critical density of the Universe at $z = 0$. The TNG100 result is to be understood as a lower limit because it does not include the \hi and \htwo in galaxies that were not included in our sample selection (Section~\ref{sec:sim:selection}). For this reason, we do not show TNG300 because the much higher stellar and gas mas limits would mean that significant gas reservoirs would be excluded. Furthermore, our calculation does not include any neutral gas not bound to haloes and subhaloes, although such gas should be negligible because virtually all non-bound gas is ionized by the UV background (hereafter UVB). We assess the importance of these effects by comparing to the $\omegahi$ calculation of \citet{villaescusanavarro_18}. While their method used a simpler prescription for the \htwo fraction, it included all gas cells in the simulation. The small difference to our galaxy sample indicates that the vast majority of \hi resides within or near galaxies, at least below $z \approx 2$. We caution that the simulation results are not fully converged, especially at high redshift (Fig. 26 in \citealt{villaescusanavarro_18}). Moreover, the \hiht separation becomes less certain at high $z$ because all our models were calibrated at low $z$, as evident from their increasing dispersion.

As this paper focuses on observational comparisons at $z = 0$, we are most interested in low-redshift constraints on $\omegahi$ and $\omegaht$. We assume that the observational data represent the total amount of \hi and \htwo in the Universe, with no severe effects due to explicit or implicit selection functions \citep[e.g.,][]{obreschkow_13}. As shown in the top panel of Fig.~\ref{fig:omega}, observations agree that $\omegahi$ falls to about $4 \times 10^{-4}$ at $z = 0$, about half of the \hi we find in TNG100 (regardless of whether galaxies or all gas are used). Thus, we expect the \hi mass function and gas fractions to also overestimate observations. For \htwo, the picture is similar (bottom panel of Fig.~\ref{fig:omega}). Here we expect virtually no molecular gas outside of galaxies \citep[see also][]{lagos_15}, though some galaxies might fall below our mass selection thresholds. Moreover, $\omegaht$ depends more strongly on the \hiht model than $\omegahi$. Despite these caveats, the \htwo abundance in TNG100 appears to be somewhat higher than observed. However, some \hiht models fall within the observational uncertainty, especially given the additional systematic uncertainties due to the CO-to-\htwo conversion factor and due to sample selection.

The redshift evolution of $\omegahi$ is characterised by relatively weak changes between $z = 4$ and $z \approx 1$. Given the large uncertainties on the data, this result does not present a conflict with observations, particularly since we have not taken into account potential selection effects. At $z < 0.5$, however, $\omegahi$ is observed to sharply decline, whereas the \hi abundance in TNG100 increases. In contrast, $\omegaht$ displays an evolution of a factor of $\approx 2$--$3$. Its peak at $z \approx 2$ is close to the peak of the cosmic SFR density \citep[e.g.,][]{madau_14}, consistent with a picture where the differences between $\omegahi$ and the cosmic SFR are owed to a changing \hi-to-\htwo conversion efficiency, driven primarily by a combination of gas column density and metallicity \citep[e.g.,][]{obreschkow_09a, lagos_14}. The bottom panel of Fig.~\ref{fig:omega} also shows the evolution of $\omegaht$ in EAGLE according to \citet{lagos_15}. Their \htwo abundance rises to a higher level around $z = 1$ but falls to a slightly lower level at $z = 0$. With rapidly improving CO data at high redshift, this evolution will be a useful constraint on galaxy formation models. In Fig.~\ref{fig:omega}, we have omitted high-$z$ CO surveys because their small volume and their selection functions complicate the interpretation. We refer the reader to \citet{popping_19} for such a comparison.

In summary, TNG100 seems to exhibit a reasonable redshift evolution of $\omegahi$ and $\omegaht$, but contains about twice as much neutral gas as observed at $z = 0$. We further discuss this tension in Section~\ref{sec:discussion:excess}.

\subsection{Mass Functions}
\label{sec:results:mfunc}

\begin{figure*}
\centering
\includegraphics[trim =  3mm 11mm 0mm 2mm, clip, scale=0.73]{\figdir/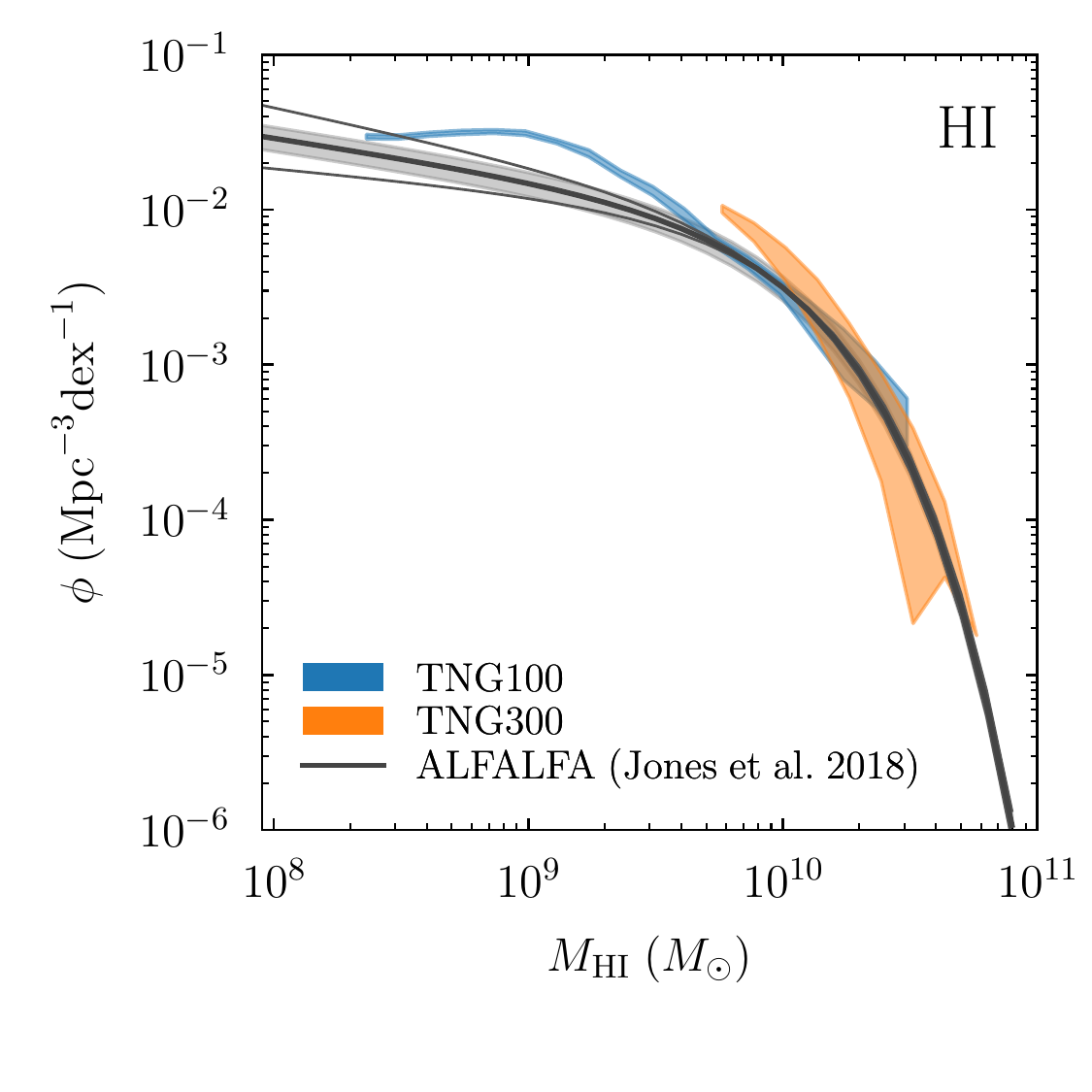}
\includegraphics[trim = 24mm 11mm 2mm 2mm, clip, scale=0.73]{\figdir/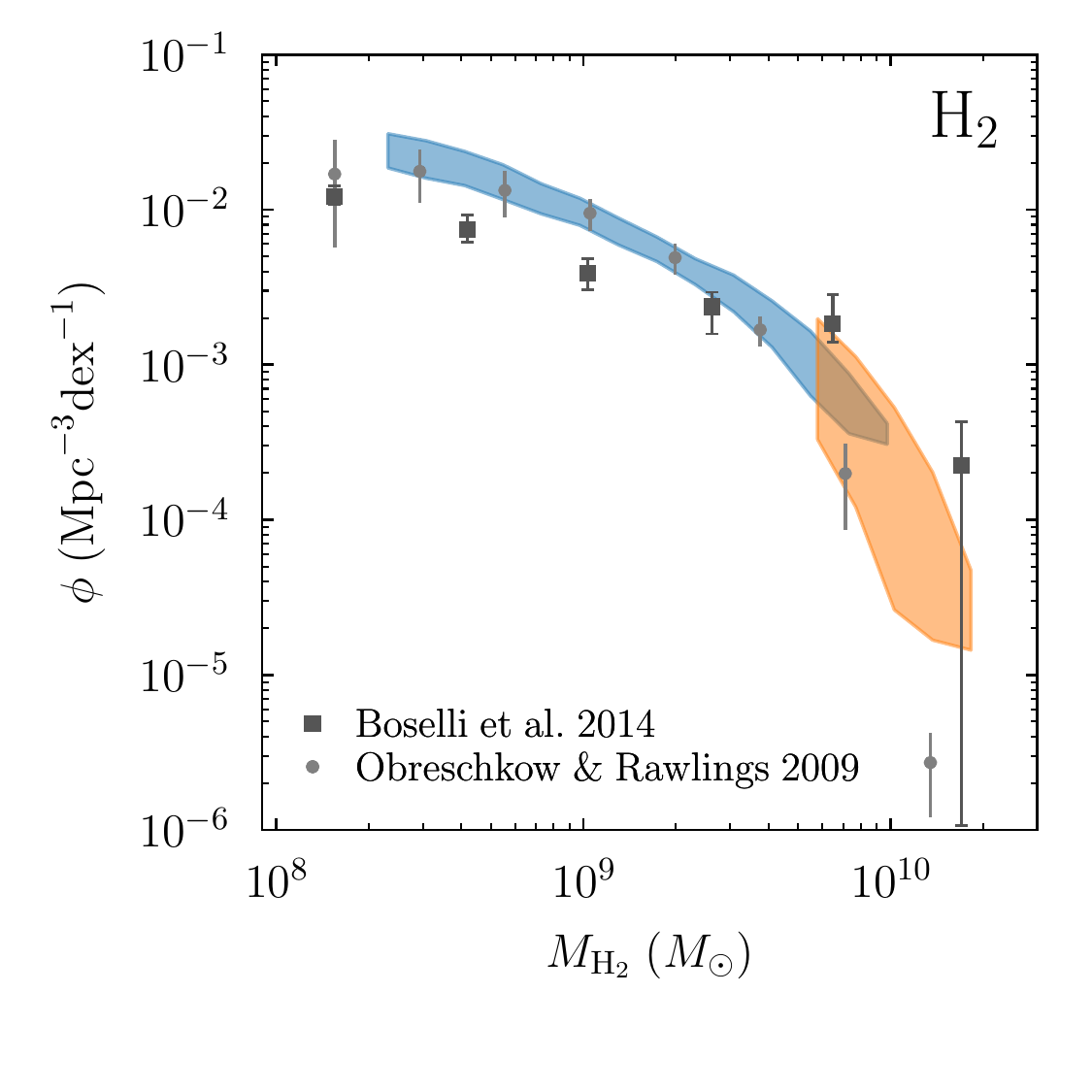}
\caption{Mass functions of atomic (left) and molecular (right) hydrogen at $z = 0$. The gray lines and points correspond to various observational measurements (Section~\ref{sec:obs:mfunc}), namely the ALFALFA \hi mass function of \citet{jones_18} and the \htwo mass function of \citet{boselli_14_hrs2}. The simulation data were mock-observed to match their respective apertures. The \htwo data of \citet{obreschkow_09a} are shown only for comparison. TNG100 and TNG300 are relatively well converged and generally match the data. At low masses, however, the \hi mass function is overestimated by a factor 2--3. Given the 1-$\sigma$ observational uncertainties in the normalization and low-mass slope (shown as a shaded area and as thin gray lines, respectively), the disagreement is statistically significant. Any disagreement with the observed \htwo mass functions is not significant given the systematic uncertainties in our modelling and in the CO-to-\htwo conversion.}
\label{fig:mfunc}
\end{figure*}

Fig.~\ref{fig:mfunc} shows the \hi and \htwo mass functions from TNG100 and TNG300 as well as a number of observational constraints. Overall, the simulations match the data fairly well, especially at the high-mass end. The \hi mass function is compared to the extremely well-measured mass function from the ALFALFA survey \citep{jones_18}. The shaded uncertainty region around the ALFALFA mass function represents the systematic error on their normalization due to the uncertainty in the overall flux calibration, the thin lines show the mass function with low-mass slopes that are one $\sigma$ lower and higher than the fiducial mass function. We cannot combine these uncertainties because \citet{jones_18} do not provide covariances. There is no discernible tension above $\mhi \gsim 5 \times 10^9 \msun$, and the TNG100 and TNG300 results agree reasonably well. Between masses of about $3 \times 10^8 \msun$ and $5 \times 10^9 \msun$, however, IllustrisTNG contains about $2$--$3$ times too many objects. This prediction is robust to the choice of \hiht model because the \htwo fraction is low, meaning that all \hiht models basically predict the same amount of \hi. Moreover, the choice of aperture has very little impact on \hi and on this mass range in particular (Fig.~\ref{fig:gasfrac_aperture}). Given the results shown in Fig.~\ref{fig:omega}, the disagreement is hardly surprising: since $\omegahi$ exceeds the $z = 0$ measurement from ALFALFA \citep{obuljen_18}, the mass function must also exceed its ALFALFA counterpart at some masses.

\citet{crain_17} showed that a similar excess in the EAGLE \hi mass function is largely due to an unphysical, resolution-dependent population of objects that contain very little dark matter or stars. We have checked that this is not the case in IllustrisTNG, i.e., that the galaxies around $\mhi \approx 10^9 \msun$ do not exhibit any obvious unphysical characteristics. Comparing to a lower-resolution version of TNG100, however, we do find a similar excess in the \hi mass function that is shifted to higher masses. A comparison with the upcoming TNG50 simulation \citep{nelson_19_outflows, pillepich_19_tng50} will reveal whether residual resolution effects change the TNG100 \hi mass function.

While the \hi mass function is more tightly constrained observationally, the \htwo mass function provides a more interesting point of comparison for our simulation because \htwo is a smaller fraction of neutral gas, meaning that the form of the \htwo mass function is less directly tied to the overall neutral gas content. The right panel of Fig.~\ref{fig:mfunc} compares the \htwo mass functions in TNG100 and TNG300 to the observations of \citet{obreschkow_09a} and \citet{boselli_14_hrs2}. The aperture chosen for the simulated galaxies was matched to the \citet{boselli_14_hrs2} observations (Section~\ref{sec:sim:apertures:h2}), the \citet{obreschkow_09a} points are shown to highlight the systematic observational uncertainty on the \htwo mass function. As with the \hi mass function, TNG100 and TNG300 are well converged and agree with the observations at the high-mass and low-mass ends. Any apparent disagreements with the observed data are not truly significant, given that our \hiht modelling is uncertain to at least a factor of two \citepalias{diemer_18_hih2} and that the observations rely on an uncertain CO-to-\htwo conversion factor \citep[see also][]{popping_19}.

\subsection{Gas Fractions}
\label{sec:results:fraction}

\begin{figure*}
\centering
\includegraphics[trim =  1mm 9mm 0mm 0mm, clip, scale=0.68]{\figdir/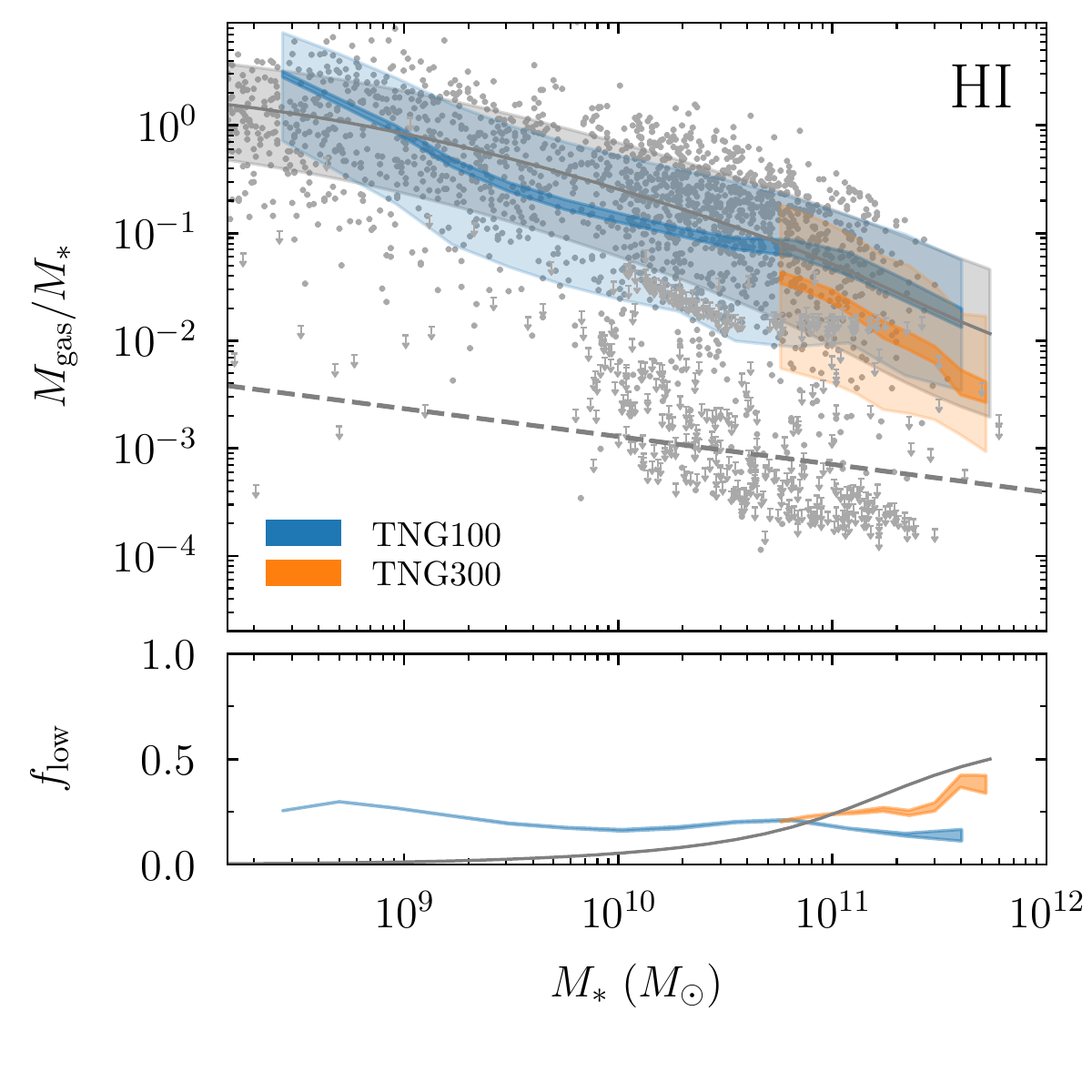}
\includegraphics[trim = 24mm 9mm 1mm 0mm, clip, scale=0.68]{\figdir/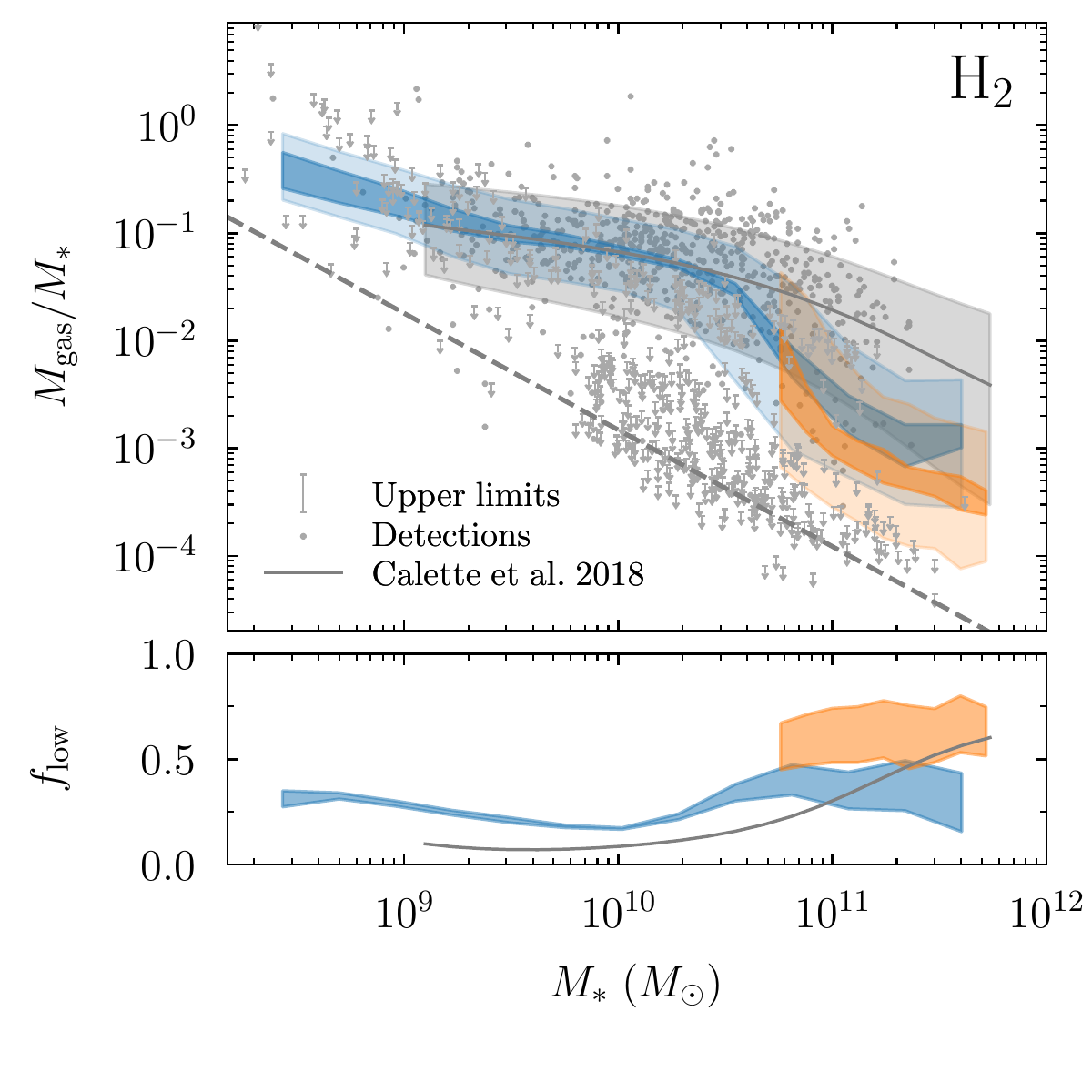}
\caption{Gas fractions of atomic (left) and molecular hydrogen (right) as a function of stellar mass at $z = 0$. The gray lines and shaded areas show the inferred median gas fractions and 68\% scatter according to the compilation of \citetalias{calette_18}. These values refer only to galaxies whose gas fractions lie above the dashed lines. The distributions of lower gas fractions could not be reliably determined in \citetalias{calette_18}, their fractional contribution is shown in the bottom panels. As a result, the median gas fractions in the top panels can appear different from those shown in \citetalias{calette_18}, especially at high stellar masses where many galaxies fall below the threshold. The simulation data are analysed in the same way, i.e., by separately counting galaxies that fall below the lower limits. The darker shaded areas show the area covered by all nine \hiht models used and can be interpreted as a systematic uncertainty. The lighter shaded area indicates the mean 68\% scatter, i.e., the mean of the scatter contours according to the different models. The detections and upper limits used in the \citetalias{calette_18} analysis are shown as gray dots and arrows (combining both their LTG and ETG samples). See Section~\ref{sec:results:fraction} for a detailed discussion of these results.}
\label{fig:fraction}
\end{figure*}

We now turn to the connection between the stellar and gaseous content of galaxies in IllustrisTNG, quantified by the fractions of \hi and \htwo mass with respect to stellar mass (hereafter simply referred to as ``gas fractions'' or $f_{\rm gas}$). Fig.~\ref{fig:fraction} shows the gas fractions of \hi and \htwo as a function of stellar mass. Only stellar mass bins with at least $50$ galaxies are shown. When comparing gas fractions, simple means or medians are not appropriate measures of the respective distributions because a number of galaxies will contain zero gas cells, or at least lie below some observational detection threshold (for example, the sensitivity limit of radio telescopes). For this reason, we separately consider two quantities: the fraction of galaxies below some threshold gas fraction, and the distribution of gas fractions above this limit. 

In defining the threshold gas fraction, we follow \citetalias{calette_18} who introduce a mass-dependent lower limit to the distributions for ETGs (dashed lines in Fig.~\ref{fig:fraction}). As we are combining the ETG and LTG samples in Fig.~\ref{fig:fraction}, we apply the same threshold to both ETGs and LTGs. \citetalias{calette_18} infer the distribution of gas fractions using Kaplan-Meier estimation, taking into account upper limits, and parameterize their results as separate fitting functions for the distribution of LTG and ETG gas fractions. We reconstruct the distribution of the full sample by summing the LTG and ETG contributions given the ETG fraction at each stellar mass (Fig.~\ref{fig:etgltg}). For both the simulated and observed distributions, we now count the fraction of galaxies that lie below the threshold, $f_{\rm low}$, and compute the median and 68\% scatter of those gas fractions that lie above the threshold. The exact value of the threshold is unimportant because it is applied to both simulation and observations. We note that $f_{\rm low}$ is not the fraction of galaxies without a gas detection, but rather the fraction of galaxies whose gas fraction falls below the threshold that is imposed a posteriori. Thus, we do not need to worry about whether a simulated galaxy would or would not have been detected by a given observational survey. If its gas fraction lies below the threshold, it is registered as such by \citetalias{calette_18}. The individual observed data points shown in Fig.~\ref{fig:fraction} highlight the importance of applying an estimator that takes into account upper limits and selection effects.

We first consider the distribution of $f_{\rm gas}$ above the threshold. Overall, the \hi fraction matches the results of \citetalias{calette_18} well, including a realistic 68\% scatter. The differences between the \hiht models are relatively small for \hi. While the median dips below the observed median by a factor of about two at intermediate stellar masses, it is unclear whether this difference is significant. TNG300 exhibits systematically lower gas abundances than TNG100. This undesirable resolution dependence could be resolved with a rescaling procedure \citep[see, e.g.,][for stellar masses]{pillepich_18}, but we do not undertake such an exercise here.

The median \htwo fraction is almost perfectly matched at stellar masses below $2 \times 10^{10} \msun$, at the highest masses it falls below the observations by a factor of about $4$--$10$ depending on the \hiht model. However, it is not clear how significant this difference is for a number of reasons. First, as discussed in Section~\ref{sec:obs:omega}, the observations are systematically uncertain due to the poorly known CO-to-\htwo conversion factor. Second, the low density of observations above $2 \times 10^{11} \msun$ means that the distribution inferred by \citetalias{calette_18} is less reliable than at lower masses. Finally, the estimated \htwo fraction becomes more model-dependent at high stellar masses (see also \citetalias{diemer_18_hih2}). Part of this effect is due to the different spatial distribution predicted by the models, which, given a limited observational aperture, can cut out significant fractions of the total \htwo mass and leads to the sharp drop in the \htwo fraction at high masses. In summary, there may be some tension between the observed and simulated \htwo fractions at the high-mass end, but it is hard to quantify this tension exactly.

We now turn to the fraction of galaxies with $f_{\rm gas}$ below the lower limit. According to \citetalias{calette_18}, this fraction increases with stellar mass, reaching about $50\%$ of both the \hi and \htwo distributions at the highest stellar masses (bottom panels of Fig.~\ref{fig:fraction}). The IllustrisTNG galaxies do not follow this trend. Especially in the \hi distribution, the fraction of low-gas galaxies decreases slightly at high masses. Most notably, the simulations contain a population of at least 25\% below the limit at all stellar masses. Virtually all of those, about 85--90\% depending on the \hiht model, are satellites. A small fraction of those turn out to be satellites that contain no gas cells at all, presumably because they were stripped away. The high satellite fraction in the low-gas galaxies can perhaps partially explain the disagreement with \citetalias{calette_18} because some observational samples are likely biased against close satellite pairs due to effects such as fibre collisions (Section~\ref{sec:discussion:gasfree}). However, in Section~\ref{sec:discussion:gasfree}, we discuss this issue further and conclude that fibre collisions cannot explain the bulk of the excess of galaxies with low gas fractions. Similarly, blending can assign some \hi mass to gas-free galaxies, but this effect impacts only a few percent of our galaxies (Appendix~\ref{sec:app:apertures}). 

We conclude that the gas fractions in IllustrisTNG broadly match the observations, but that there is an excess of satellites with very low gas fractions. By considering LTGs and ETGs as one combined sample, we have sidestepped the question of how the gas fractions correlate with the morphological type of galaxies; we return to this question in Section~\ref{sec:results:morphology}.

\subsection{Correlations with Star Formation Rate}
\label{sec:results:sfr}

\begin{figure}
\centering
\includegraphics[trim = 7mm 28mm 6mm 2mm, clip, scale=0.72]{\figdir/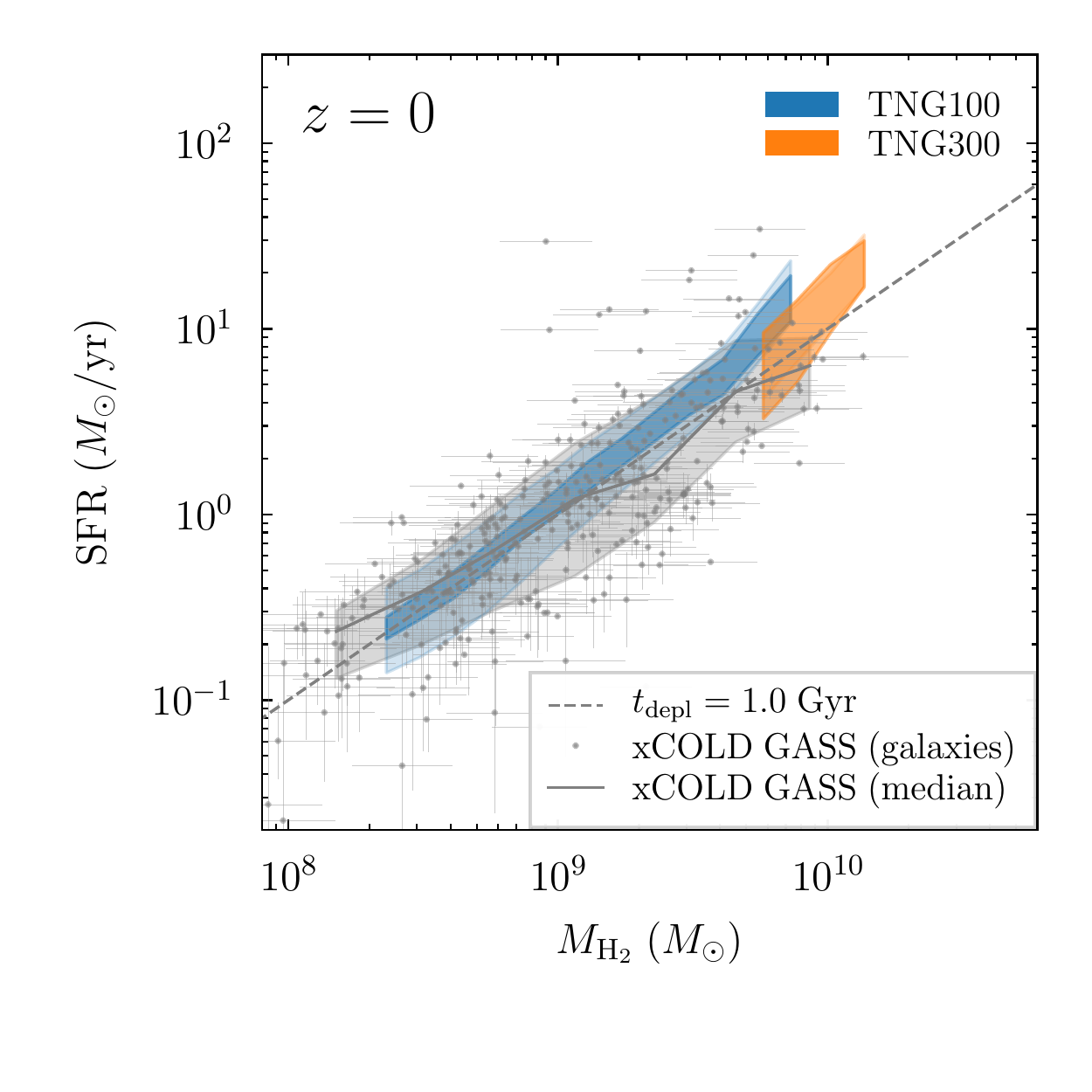}
\includegraphics[trim = 7mm 13mm 6mm 5mm, clip, scale=0.72]{\figdir/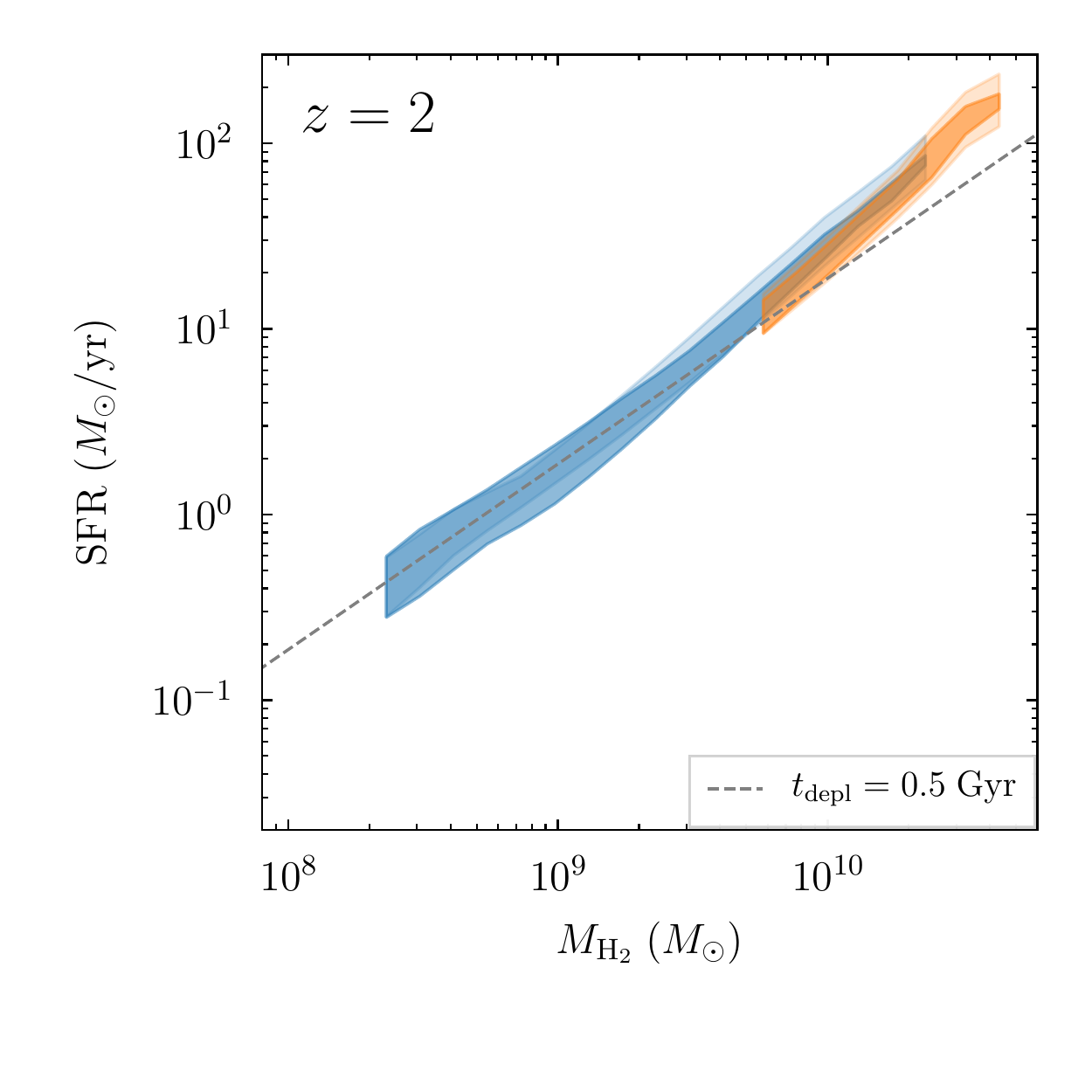}
\caption{Relation between molecular gas mass and SFR at $z = 0$ (top) and $z = 2$ (bottom). In the top panel, the gray points show individual galaxies from xCOLD GASS \citep{saintonge_17}, the solid line and shaded region show their median and 68\% scatter. IllustrisTNG matches the median relation well, to better than a factor of two at all masses. In TNG300, the scatter is smaller than the variation due to the \hiht models. The relation closely approximates a constant depletion time as parameterized by \citet[][dashed gray lines]{tacconi_18}.}
\label{fig:sfr}
\end{figure}

Given the reasonable gas fractions and the generally realistic star formation activity in IllustrisTNG \citep{donnari_19}, we expect a that the molecular gas reservoirs of galaxies should tightly correlate with their SFR. This trend is well established observationally (Fig.~\ref{fig:sfr}). We note that the $\mht$-SFR correlation is not quite the same as the Kennicutt-Schmidt relation \citep{schmidt_59, kennicutt_98} because we have not normalised the mass and SFR by area.

The top panel of Fig.~\ref{fig:sfr} shows the median SFR as a function of $\mht$ for both observations and simulations at $z = 0$. The xCOLD GASS galaxy sample is representative in the sense that it was randomly selected from a particular stellar mass range \citep{saintonge_12, saintonge_17}, and that galaxies were observed until a certain \htwo fraction was detected or a certain upper limit was reached. In particular, galaxies in the COLD GASS sample ($\mstar > 10^{10} \msun$) were guaranteed to be detected in CO if $\mht / \mstar > 1.5\%$, while galaxies in the lower-mass COLD GASS-low sample ($10^9 < \mstar < 10^{10} \msun$) were detected if $\mht / \mstar > 2.5\%$. To match this selection, we have chosen all simulated galaxies in the respective stellar mass ranges that reach the limiting \htwo mass.  Given the mixed sources of the SFR measurements, there is no clear detection limit on the SFR \citep{saintonge_17}. We take the lowest SFR in the sample, $10^{-2.8} \msun/{\rm yr}$, as a guideline and neglect all IllustrisTNG galaxies whose SFR falls below this value. In practice, this cut makes no discernible difference. 

The median trend at $z = 0$ is matched well by IllustrisTNG, to a factor of two or better at all masses. For TNG300, the situation is less clear because the data become somewhat sparse at the highest masses. The scatter in TNG100 appears to be slightly smaller than observed, but the observed scatter constitutes an upper limit because it does not take into account the sizeable error bars on the individual SFR measurements. Moreover, the observed SFRs correspond to an average over a few hundred Myr and might therefore be subject to added scatter compared to the instantaneous SFRs used for the simulation data (Section~\ref{sec:sim:apertures:sfr}).

An alternative measure by which to understand the relation between gas mass and star formation is the depletion time, $t_{\rm depl} \equiv \mht / {\rm SFR}$, i.e., the time over which a gas reservoir would be exhausted given the current SFR. Observationally, the depletion time is thought to be more or less independent of mass but to evolve with redshift. The dashed lines in Fig.~\ref{fig:sfr} correspond to a constant depletion time as parameterized by \citet{tacconi_18}, $t_{\rm depl} \approx 1\ {\rm Gyr} \times (1 + z)^{-0.57}$. IllustrisTNG reproduces this relation between $z = 0$ and $z = 2$, although the highest-mass galaxies have higher star formation rates than expected by a factor of about two. The comparison to \citet{tacconi_18} should be taken as approximate because we have not matched our galaxy selection to theirs, which includes only galaxies on the star-forming main sequence. However, we have checked that the $\mht$--SFR connection does not significantly differ between ETGs and LTGs (defined as described in Section~\ref{sec:sim:morphology}). 

\citet{lagos_15} reported a similar result for the EAGLE simulation, namely a roughly constant depletion time of $t_{\rm depl} \approx 10^9$ Gyr at $z = 0$. They argued that an overly tight $\mht$--SFR relation is perhaps to be expected because the star formation in simulations is directly based on the density of gas, and thus the density of cold gas in the ISM models. 

\subsection{Spatial Distribution}
\label{sec:results:size}

Having investigated how atomic and molecular gas is distributed between galaxies of different stellar mass and SFR, we now turn to the distribution of gas within galaxies. 

\subsubsection{The \hi Mass-size Relation}
\label{sec:results:size:hi}

\begin{figure}
\centering
\includegraphics[trim = 5mm 8mm 6mm 2mm, clip, scale=0.72]{\figdir/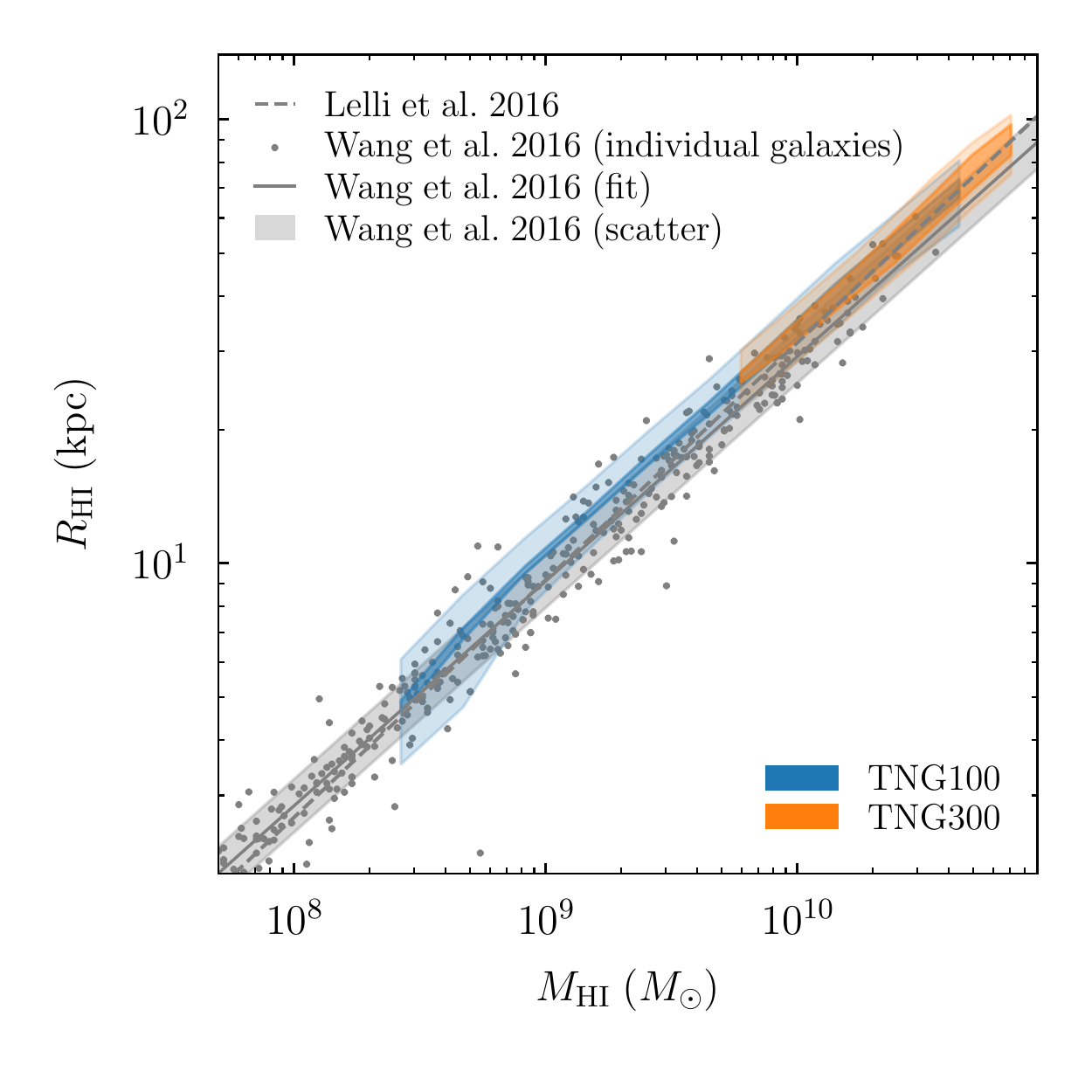}
\caption{The \hi mass-size relation at $z = 0$, where $\rhi$ is defined as the radius where $\sigmahi$ falls below $1 \msun/{\rm pc}^2$. The gray points show the data compilation of \citet{wang_16_hi}, the solid gray lines shows their best linear fit. For comparison, the dashed line shows the fit from \citet{lelli_16}.}
\label{fig:hi_mass_size}
\end{figure}

Fig.~\ref{fig:hi_mass_size} shows the \hi mass-size relation as measured in TNG100 and TNG300, and in the observational compilations of \citet{wang_16_hi} and \citet{lelli_16}. Observationally, the relation is extremely tight, with only $0.06$ dex scatter (15\%, shaded gray). Its slope is closed to $0.5$, the value one would obtain for a disc of constant surface density ($0.506$ in \citealt{wang_16_hi}, 0.535 in \citealt{lelli_16}). Both TNG100 and TNG300 match the slope of the relation almost perfectly, but predict very slightly larger \hi radii than observed, about 14\% above the measured relation. The \hiht models agree to better than 13\% for all bins for TNG100 and to better than 20\% for TNG300, a much more well-defined prediction than for gas fractions or mass functions. Interestingly, TNG100 and TNG300 are almost perfectly converged in this metric, unlike in any other quantity we consider. We discuss the physical reasons for the tight relation and the good agreement in Section~\ref{sec:discussion:hisize}.

\subsubsection{\hi Surface Density Profiles}
\label{sec:results:size:hiprofiles}

\begin{figure}
\centering
\includegraphics[trim = 2mm 24mm 6mm 2mm, clip, scale=0.69]{\figdir/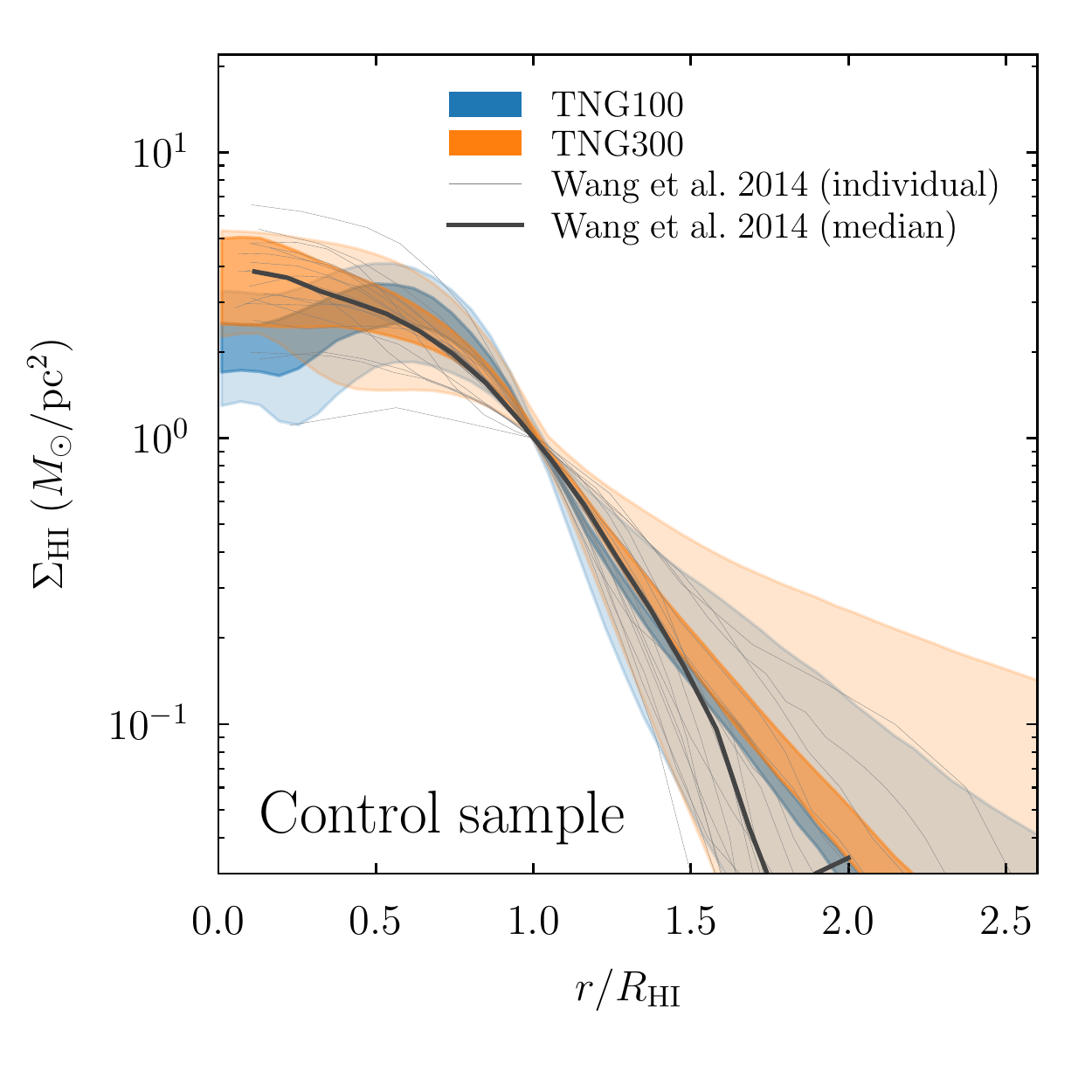}
\includegraphics[trim = 2mm  8mm 6mm 3mm, clip, scale=0.69]{\figdir/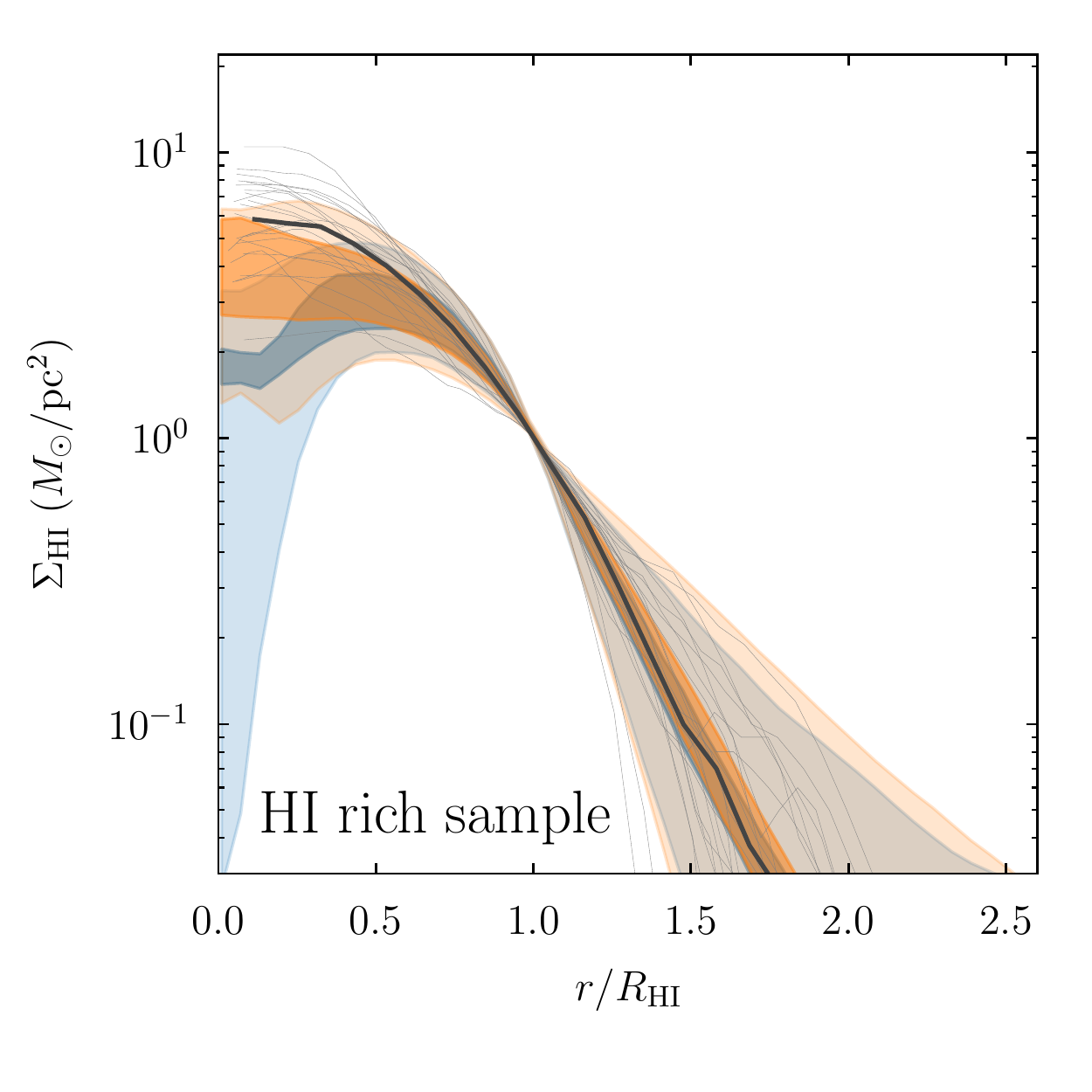}
\caption{Median radial profiles of the \hi surface density. The light shaded areas show the 68\% scatter. As the profiles are scaled to $\rhi$, they overlap at $\sigmahi = 1 \msun/{\rm pc^2}$ by definition. The gray lines show individual galaxies and the median for the Bluedisk sample of \citet{wang_14_hi}. The top panel shows their control sample ($10^{9.1} < \mhi < 10^{9.8} \msun$), the bottom panel their \hi-rich sample ($\mhi > 10^{9.8} \msun$).}
\label{fig:prof_hi}
\end{figure}

Having established that $\rhi$ matches observations well, we proceed to the radial distribution of \hi. Fig.~\ref{fig:prof_hi} shows the radial \hi profiles from the Bluedisk survey \citep{wang_13_bluedisks, wang_14_hi}. Their sample was designed to contain an \hi-rich and a control subset (Section~\ref{sec:obs:sizes}). To mimic their selection, we choose all galaxies with $10^{10} < \mstar < 10^{11} \msun$, $0.5 < {\rm SFR} < 10\ \msun/{\rm yr}$, and a stellar half-mass radius of at least $3$ kpc (though the latter cut makes virtually no difference). For the control sample, we select galaxies with $10^{9.1} \msun < \mhi < 10^{9.8} \msun$, for the \hi-rich sample we apply a minimum of $\mhi > 10^{9.8} \msun$ (according to the respective \hiht model). \citet{bahe_16} carefully demonstrated that these relatively simple cuts match the Bluedisk selection closely.

Observationally, the median \hi profiles for the two samples turn out to be almost indistinguishable, but the \hi-rich IllustrisTNG galaxies do exhibit somewhat different profiles from the control galaxies. The median profiles of the control sample match the data reasonably well, though TNG100 galaxies exhibit a slight deficit at small radii. In both TNG100 and TNG300, the \hi profiles in the control sample are perhaps a little too extended at large radii, though the difference is well within the scatter of the observations. In the more massive \hi-rich sample, the median outer profiles are matched extremely well but the simulations show clear evidence of central \hi holes: up to a factor of five deficit of \hi at the centre. This trend is more pronounced in TNG100, but both TNG100 and TNG300 show large scatter at the smallest radii. The overall features of the profiles are independent of the \hiht model, which is expected at large radii where \hi dominates over \htwo, but not necessarily at small radii where the molecular fraction could be high.

\hi holes have also been observed in some galaxies \citep[e.g.][]{boomsma_08} but are not prevalent in the Bluedisk sample. We note that the holes in IllustrisTNG galaxies are not likely to be a resolution effect. First, they are more pronounced in the higher-resolution TNG100 than in TNG300. Second, they are more pronounced in the \hi-rich sample, which contains the more massive, and thus more extended, \hi disks. According to the median \hi mass-size relation, $\rhi \approx 30$ kpc at the lower-mass end of this sample, about $40$ times the force resolution length in TNG100. Finally, the holes are not unique to either volumetric or projected models but a robust prediction of both. 

The question, then, arises as to what causes the central \hi holes. There are two possible processes that can create them. First, gas at the center can be ionized or ejected, for example due to feedback from active galactic nuclei (AGN) in large galaxies \citep{weinberger_18}. Second, the inner part of the gas disc can have high enough surface densities to almost entirely transition to \htwo. We find evidence for the former mechanism \citep[see also][]{nelson_19_outflows}. First, the prevalence of \hi holes increases with mass, a trend that would be expected for AGN feedback. Second, the median neutral fraction, $\fneutral$, drops towards the centre of massive galaxies, indicating that the majority of the gas is ionized in many objects. Third, the prevalence of \hi holes depends on the star formation activity: galaxies with a specific SFR (sSFR) above $10^{-10}/{\rm yr}$ exhibit barely any \hi holes, whereas the vast majority of galaxies below an sSFR of $10^{-11}/{\rm yr}$ do exhibit them. Thus, \hi holes do not seem to be connected to a dominance of \htwo over \hi. The control and \hi-rich samples select slightly different SFRs, explaining the different median profiles in Fig.~\ref{fig:prof_hi}. 

\citet{bahe_16} analysed \hi profiles in the EAGLE simulation and report that a significant fraction of their galaxies exhibit clear \hi holes, with holes becoming more prevalent towards high \hi masses (their figure 5). The holes in EAGLE, however, appear not only at the center but also throughout the disc, and are traced to their particular feedback implementation \citep{bahe_16}.

\subsubsection{The Extent of the \htwo Distribution}
\label{sec:results:size:h2}

\begin{figure}
\centering
\includegraphics[trim = 2mm 8mm 4mm 2mm, clip, scale=0.69]{\figdir/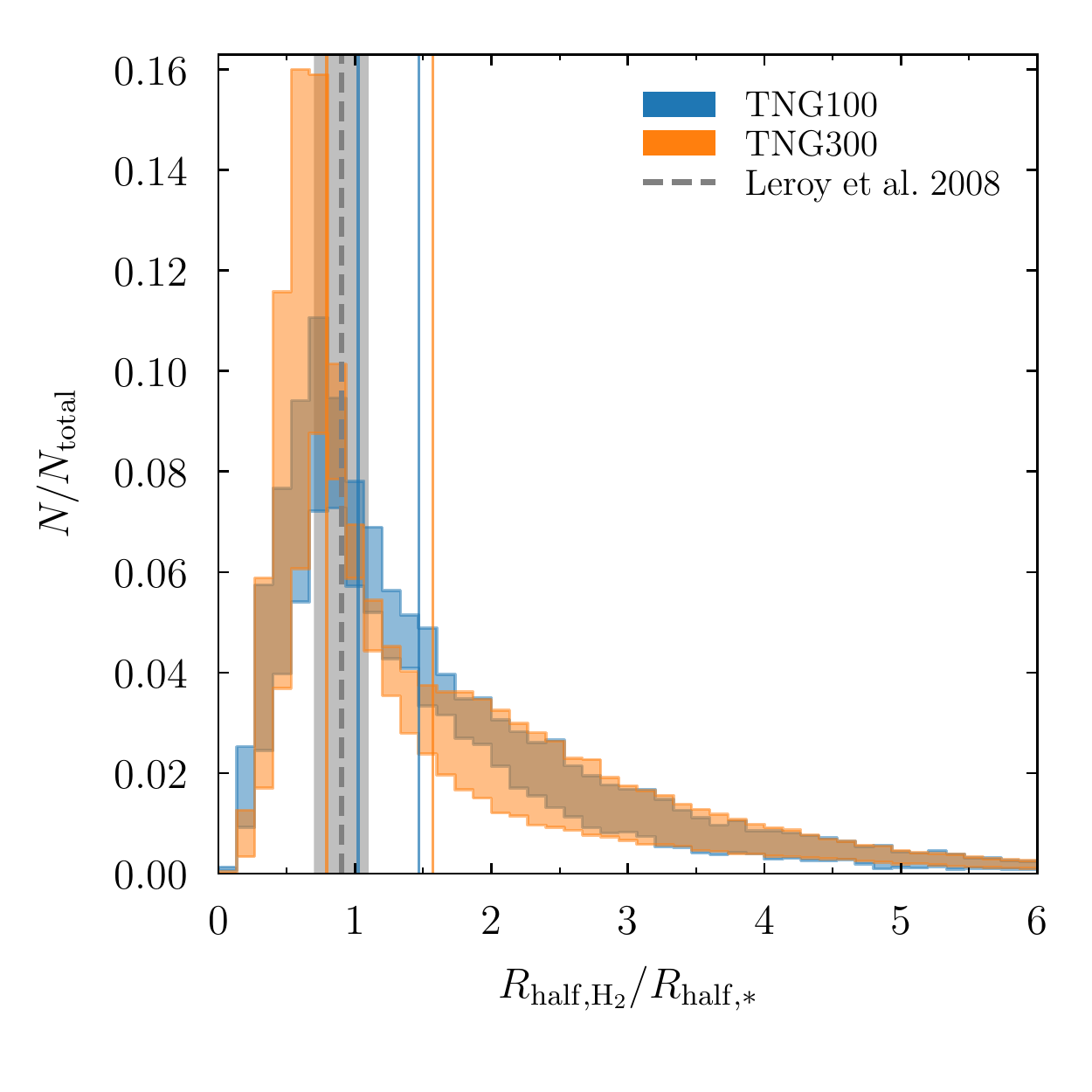}
\caption{Distribution of the ratio of \htwo and stellar half-mass radii, a crude test of the extent of \htwo compared to the optical size of the galaxy. The gray histogram shows the distribution of CO-to-stellar half-mass radii from the EDGE-CALIFA survey \citep{bolatto_17}. Their best-fit CO-to-stellar scale ratio is unity (dashed line), whereas IllustrisTNG galaxies exhibit somewhat larger median ratios and large scatter (the thin vertical lines bracket the medians according to the different \hiht models).}
\label{fig:h2_size}
\end{figure}

\begin{figure*}
\centering
\includegraphics[trim =  1mm 9mm 0mm 0mm, clip, scale=0.62]{\figdir/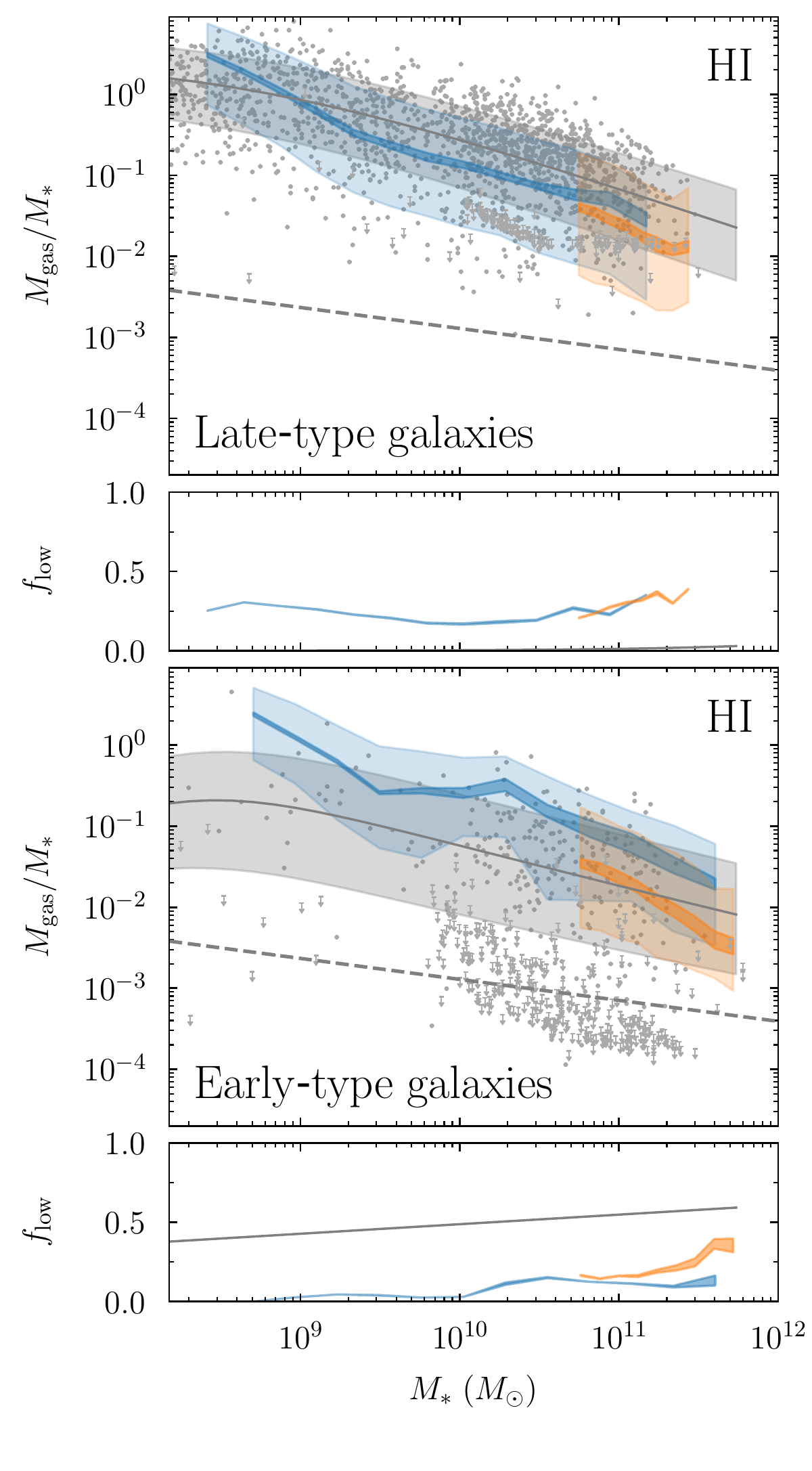}
\includegraphics[trim = 24mm 9mm 1mm 0mm, clip, scale=0.62]{\figdir/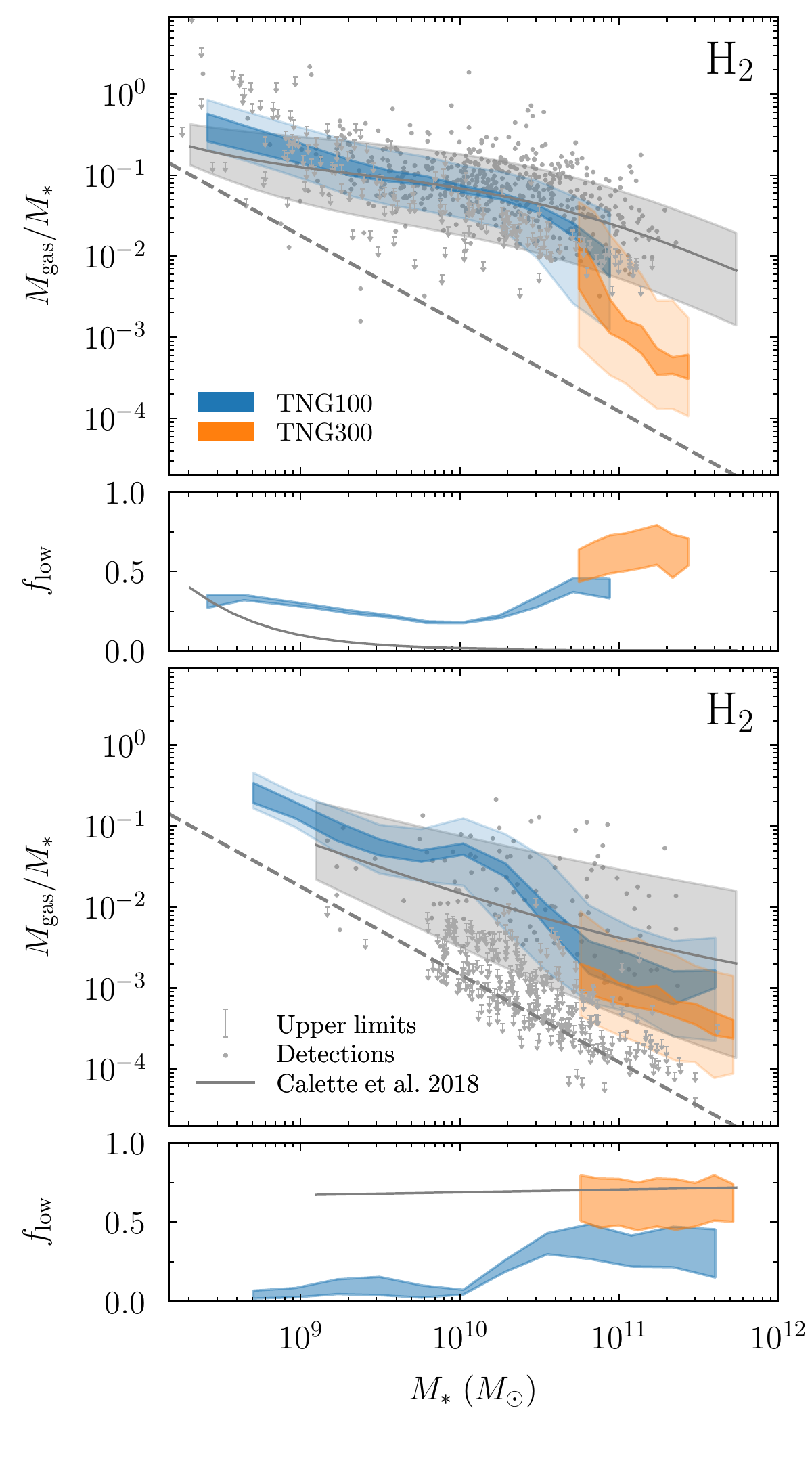}
\caption{Same as Fig.~\ref{fig:fraction}, but with the galaxy sample split into LTGs (top row) and ETGs (bottom row) by the concentration of their stellar mass. The \citetalias{calette_18} constraints are derived directly from their best-fit distributions. The match to observations is somewhat worse than for the overall sample, highlighting that the gas content of IllustrisTNG galaxies does not correlate with morphology as observed in the real Universe.}
\label{fig:gasfrac_c}
\end{figure*}

As discussed in Section~\ref{sec:obs:sizes}, there are a number of factors that complicate the interpretation of observed molecular profiles. Thus, we restrict our comparison to a simple measure of the relative extent of the molecular and stellar distributions: the ratio of \htwo and stellar half-mass radii, shown in Fig.~\ref{fig:h2_size}. The gray histogram shows the ratio of CO and stellar half-mass radii from the EDGE-CALIFA survey \citep[][see their figure 12]{bolatto_17}. The CALIFA selection function is based mostly on the optical radii of galaxies \citep{walcher_14}, but the EDGE selection complicates this picture further. To mimic the overall selection, we choose all simulated galaxies that fall within the ranges of galaxy properties observed in the final EDGE-CALIFA sample, namely, a projected stellar half-mass radius between $1.3$ and $8.8$ kpc, $10^{9.5} \msun < \mstar < 10^{12} \msun$, $\mht > 10^8 \msun$, and an SFR between $0.01$ and $100\ \msun / {\rm yr}$. We have verified that our results are not sensitive to the exact values of any of these limits.

\citet{bolatto_17} find a relatively tight relation between the CO and stellar half-mass radii, with a best-fit ratio of 1.00 (dashed line in Fig.~\ref{fig:h2_size}). In agreement with these observations, a large fraction of IllustrisTNG galaxies exhibit radius ratios around unity, but there is also a significant tail towards large ratios. These galaxies have an \htwo distribution that is much more extended than the stellar distribution, whereas no such objects were observed in EDGE-CALIFA. As a result, the median ratio is closer to $1.3$ in the simulations, though with a large dependence on the \hiht model. The median ratio tends to increase towards higher \htwo masses. The stellar sizes in IllustrisTNG match observations well \citep{genel_18}, meaning that the disagreements are likely due to the extent of the \htwo distribution. The \htwo radii appear well converged, with small differences between TNG100 and TNG300, or with the lower-resolution version TNG100-2. Splitting the sample into LTGs and ETGs as described in Section~\ref{sec:sim:morphology} has only a modest effect on the results.

While half-mass radii have the advantage of being simple to measure, observed CO and stellar profiles are often fit with an exponential \citep[e.g.,][]{leroy_08}. This procedure tends to give very similar results: when comparing the fitted scale lengths of the CO and stellar profiles, \citet{bolatto_17} find a best-fit relation of $l_{\rm CO} = 0.89 l_*$, \citet{leroy_08} find $l_{\rm CO} \approx (0.9 \pm 0.2) l_*$. We thus conclude that our comparison is robust to the exact definition of the scale radii. 

There are, however, a number of other caveats. For example, the CO-to-\htwo conversion factor could depend on radius, leading to CO profiles that would systematically differ from \htwo profiles \citep[][and references therein]{bolatto_13}. Moreover, the simple CO-to-stellar half-mass ratio shown in Fig.~\ref{fig:h2_size} may paint too simplistic a picture because the CO surface density profiles of galaxies have been found to be diverse \citep[e.g.,][]{young_91, regan_01, leroy_08, leroy_09}. For example, CO emission at large radii has been seen at high redshift \citep{cicone_15, ginolfi_17, dannerbauer_17}. While these complications mean that we cannot draw strong conclusions from Fig.~\ref{fig:h2_size}, there is another piece of evidence that our simulated \htwo distributions are too extended: we showed that IllustrisTNG does not quite match the Kennicutt-Schmidt relation at the high-surface-density end, with low $\sigmaht$ at fixed SFR (Fig.~3 in \citetalias{diemer_18_hih2}). This mismatch indicates that star formation in the simulation happens at lower surface densities than observed, which could well be a resolution effect.

In summary, we find tentative evidence that the spatial distribution of \htwo in our modelling is somewhat more extended than in observations. We leave a more detailed comparison for future work.

\subsection{Correlation with Morphology}
\label{sec:results:morphology}

When considering gas fractions as a function of stellar mass in Section~\ref{sec:results:fraction}, we combined the \citetalias{calette_18} results for LTGs and ETGs to recover the distribution of the overall galaxy sample. Considering the types separately, however, tightens the scatter on the relations and allows us to check whether the gas content of IllustrisTNG galaxies correlates with their stellar morphology as expected. To this end, Fig.~\ref{fig:gasfrac_c} shows the same comparison of gas fractions as Fig.~\ref{fig:fraction}, but split into LTGs (top row) and ETGs (bottom row) according to stellar mass concentration (Section~\ref{sec:sim:morphology} and Appendix~\ref{sec:app:morphology}).

This comparison reveals a number of issues. First, the fractions of galaxies below the $f_{\rm gas}$ threshold trends the wrong way: a significant number of LTGs have no gas when almost no such objects are observed, whereas there are not enough low-gas ETGs. Clearly, a lack of neutral gas does not correlate well with the shape of the stellar distribution as captured by concentration. Second, the median \hi fractions in the ETG sample are too high. While we found that the median gas fractions and scatter of the overall galaxy sample are in good agreement with data, we conclude that these gas fractions are not always assigned to galaxy samples with the expected morphological properties.

However, when splitting the sample by colour, we find much better agreement with the observed gas fractions than in Fig.~\ref{fig:gasfrac_c}. This conclusion is in line with the work of \citet{rodriguezgomez_18}, who found that galaxy morphology and colour in IllustrisTNG do not correlate quite as observed. In Appendix~\ref{sec:app:morphology}, we further discuss the connection between gas content, various morphological indicators, and colour. We conclude that the gas content of galaxies correlates much more strongly with colour than with morphology in IllustrisTNG.

\subsection{Results for the Original Illustris Simulation}
\label{sec:results:orig}

The IllustrisTNG physics model represents a major improvement over the original Illustris model \citep{pillepich_18_tng}. However, it is still informative to evaluate the original simulation with regard to its \hi and \htwo content. For this purpose, we provide Illustris-1 versions of Figs.~\ref{fig:omega} through \ref{fig:h2_size} at \href{http://www.benediktdiemer.com/data/}{benediktdiemer.com/data}. In this section, we briefly describe the main conclusions. We omit Fig.~\ref{fig:gasfrac_c} from the set and from our discussion because our chosen morphological indicator, $C_{80,20}$, takes on such different values in Illustris-1 that it cannot be used for a sensible morphological classification. 

Most notably, $\omegahi$ and $\omegaht$ are massively overestimated in Illustris-1 at $z = 0$, by almost an order of magnitude. This neutral gas excess also manifests itself in the \hi and \htwo mass functions, which are similarly overestimated by up to an order of magnitude at some masses. It follows that the gas fractions also exceed observations, by a factor of about five at the low-mass end. They decrease strongly towards high masses, falling below the \citetalias{calette_18} relations at $M_* > 10^{11} \msun$. This trend is also apparent in Fig.~3a of \citet{vogelsberger_14_nature}, but the \citet{huang_12} ALFALFA gas fractions shown there are somewhat higher than than inferred by \citetalias{calette_18}, leading to a seemingly better agreement at the low-mass end. Interestingly, Illustris-1 matches the fraction of galaxies below the threshold better than IllustrisTNG as it contains a much smaller population of very gas-poor objects and follows the expected trends with stellar mass. Given the excess of both \hi and \htwo, it is unsurprising that the correlation between $\mht$ and SFR at $z = 0$ is similarly off, with a large and non-constant depletion time. At $z = 2$, however, Illustris-1 does roughly match the constant depletion time found in IllustrisTNG.

Despite these significant disagreements, the \hi mass-size relation turns out to be robust once again: while slightly shallower in Illustris, the normalization is matched at $\mhi \approx 10^9 \msun$. This agreement is somewhat unexpected, given that the stellar sizes of galaxies were about twice too large below $\mstar \lsim 10^{10.5} \msun$ in Illustris-1, an observable that is much improved in TNG100 \citep{pillepich_18_tng, genel_18}. Thus, the improvement in stellar sizes was not driven by a change in the distribution of neutral gas. This conclusion is confirmed by the \hi profiles, which more or less agree between Illustris-1 and IllustrisTNG. Conversely, the median \htwo half-mass radius in Illustris-1 is significantly larger than in IllustrisTNG.

In summary, Illustris-1 galaxies contain a significant excess of neutral gas at $z = 0$, leading to offsets in all correlations with other galaxy properties. However, the properties of the excessively large \hi discs are internally consistent. Our findings highlight the enormous improvements of the IllustrisTNG galaxy formation model compared to the original model. 


\section{Discussion}
\label{sec:discussion}

In this section, we discuss the physics behind the most striking agreements and tensions with observations that we have discovered. We also consider anecdotal evidence from individual galaxies such as the Milky Way.

\subsection{Is There an Excess of Neutral Gas in IllustrisTNG?}
\label{sec:discussion:excess}

Perhaps the most significant tension with observations is that IllustrisTNG appears to contain about twice as much neutral gas at $z = 0$ as expected observationally, despite efficient stellar and AGN feedback. This conclusion is entirely independent of the \hiht modelling, which takes the neutral gas abundance as an input from the simulation. In non-star forming cells (below the density threshold of $n_{\rm H} = 0.106\ {\rm cm}^{-3}$), the neutral fraction is determined numerically by the balance between cooling, the photoionization rate due to the UVB \citep{fauchergiguere_09} and due to nearby AGN, and self-shielding according to a fitting function \citep[][see \citealt{vogelsberger_13} for details]{rahmati_13}. In cells above the density threshold, the \citet{springel_03} two-phase ISM model governs the gas physics. There, we assume that all gas in cold clouds is entirely neutral whereas the hot phase is entirely ionized, leading to a neutral fraction of about 90\%. If the fraction of star-forming gas was drastically different, we would expect corresponding changes in the stellar properties of IllustrisTNG galaxies, which more or less agree with observations. Given these considerations, and assuming that the observations are accurate, there are a number of possible reasons why the neutral fraction might be overestimated in star-forming cells, in non-star-forming cells, or in both.

In reality, some part of the cold-cloud phase in the \citet{springel_03} model could be ionized due to radiation from young stars \citep[][see also Section~4.3 of \citetalias{diemer_18_hih2}]{rahmati_13_localradiation}. However, only about 30\% of the total neutral gas in TNG100 stems from star-forming cells, meaning that even erasing all neutral gas from them would not reduce the overall neutral abundance by a factor of two.

In non-star-forming gas, the neutral fraction could be lowered if the gas was heated or experienced stronger photo-ionization. For example, stronger feedback could remove some gas that is dense enough to self-shield and contain a significant neutral component. Another explanation could be the strength of the UVB: if the model of \citet{fauchergiguere_09} underestimated the true background significantly, the neutral fraction would be artificially increased, perhaps without significantly increasing the star formation activity. We have tested this mechanism using a smaller test simulation that uses the \citet{puchwein_19} UVB instead of \citet{fauchergiguere_09}. In the test simulation, the neutral gas abundance is reduced by about 10\%, indicating that the UVB would need to change somewhat drastically to reduce the neutral gas masses by a factor of two.

While none of these effects seem likely to account for the entire disagreement in the neutral gas abundance, several of them could conspire. We note that the disagreement in $\omegahi$ is driven largely by an upturn between $z = 1$ and $z = 0$ that is not apparent in observations (Fig.~\ref{fig:omega}). The upturn is partially caused by a decrease in the \htwo abundance over the same time interval, but the combined neutral abundance does increase in TNG100. An understanding of the reasons for this trend might shed light on the disagreement in $\omegahi$, an investigation we leave for future work.

\subsection{What is the Origin of Gas-free Galaxies in IllustrisTNG?}
\label{sec:discussion:gasfree}

While investigating \hi and \htwo fractions in Section~\ref{sec:results:fraction}, we discovered that IllustrisTNG contains a significant population of satellites whose gas fractions lie below the detection threshold. Upon closer inspection, some of those galaxies turn out not to contain any gas at all, that is, the halo finder associates no gas cells with them despite their significant stellar masses. Such gas-free galaxies account for 7\% of our TNG100 sample ($M_* > 2 \times 10^8 \msun$) and are found across the entire range of stellar mass. While gas-free galaxies account only partially for the excess of galaxies below the limit in Fig.~\ref{fig:fraction}, they are, at first sight, a strange population that we wish to investigate further.

The vast majority of gas-free galaxies, 94\%, are satellites. They can occupy (sub-)halos as massive as $10^{12} \msun$, but live in dark matter halos that are, on average, about $0.5$ dex less massive than those of their gaseous counterparts. About 20\% of them contain no dark matter at all, meaning that they are purely stellar clumps. Those objects are mostly tagged as possible halo finder artefacts \citep[][see also \citealt{genel_18} and \citealt{pillepich_18}]{nelson_19_datarelease}. Conversely, out of those gas-free galaxies that do have dark halos, only 3\% are tagged. While the average satellite in our sample resides at a median radius of $0.8 \rtoc$ (where $\rtoc$ refers to the halo radius of the host halo), the gas-free satellites live closer to their hosts, at a median distance of $0.55 \rtoc$. We find virtually the same distribution of distances for the larger population of satellites that fall into the low-gas category in our comparison of gas fractions (Fig.~\ref{fig:fraction}, see also \citealt{marasco_16}). From a visual inspection of the stellar distribution around some gas-free satellites, we conclude that some of them are relatively isolated, but that their majority lives in crowded environments, as expected from their radial distribution.

These properties constitute strong evidence that the majority of gas-free galaxies have been stripped of their gas and most of their dark matter by a larger host halo \citep[see, e.g.,][for related investigations in IllustrisTNG]{yun_19_jellyfish, stevens_19_hi}. The small fraction of gas-free galaxies that are identified as centrals at $z = 0$ may well have had a significant encounter in the past. Interestingly, the stellar size distributions of gas-free galaxies are compatible with those of the overall sample. Thus, it appears that the stellar population is relatively immune to the processes that so efficiently strip dark matter and gas from galaxies \citep[cf.][]{niemiec_19}, presumably because it is much more concentrated than the dark and gaseous components. 

In terms of our comparison to observations, the question is whether low-gas satellites would be included in the observational samples used in the \citetalias{calette_18} compilation. The answer depends somewhat on whether we consider \hi or \htwo, and on the stellar mass range. At low masses, where we see a strong excess in the fraction of galaxies with low \hi mass, the \hi fractions used in \citetalias{calette_18} are dominated by the Updated Nearby Galaxy Catalogue of \citet{karachentsev_13}, which explicitly includes objects in crowded environments. For the ALFALFA-based measurements, we expect no explicit bias against satellites as ALFALFA is a blind survey. The only considerable bias against low-gas satellites could stem from blending, where satellites are assigned additional \hi mass due to confusion with other sources (most likely, their central galaxy). However, we show in Appendix~\ref{sec:app:apertures} that blending is a relatively minor effect in our sample. 

As there are currently no blind CO surveys at low redshift, many of the \htwo fraction samples are based on an optical selection using SDSS, meaning that those samples might be biased against close pairs due to fibre collisions \citep{zehavi_02}. However, the median distance of satellites to their host in our TNG100 sample is 140 kpc, which would correspond to the 55'' fibre collision limit at a redshift of $z \approx 0.14$. Since samples such as GASS and COLD GASS have much lower median redshifts, the bulk of their satellite population should not be affected, and there are no further cuts on the proximity to another galaxy \citep{catinella_10, catinella_13, saintonge_11_coldgass, saintonge_17}. Thus, the impact of fibre collisions remains limited.

In conclusion, IllustrisTNG hosts a population of satellites with very little or no neutral gas, and there is no clear reason why the bulk of such galaxies should not be included in observational samples. Excessive stripping is not likely to be responsible, as \citet{stevens_19_hi} showed that the reduction in \hi mass in TNG100 is compatible with observations \citep{brown_17}. We conclude that, despite the overall excess in neutral gas discussed in the previous section, too many galaxies in IllustrisTNG are left with small amounts of neutral gas, perhaps due to excessive feedback. Such issues provide motivation to push observations of the \hi content of galaxies to lower gas fractions, for example using new telescopes such as FAST \citep{nan_11}.

\subsection{What Sets the \hi Mass-size Relation?}
\label{sec:discussion:hisize}

In Section~\ref{sec:results:size:hi}, we found that IllustrisTNG matches the observed \hi mass-size relation with surprising accuracy. The \hiht models agree almost exactly, an indication that the \hi radius is generally not set by the \hiht transition (see \citet{wang_16_hi} for observational arguments to the same effect). Instead, one could imagine that \hi sizes were driven by the transition from ionized to neutral gas, and thus by the interplay of the UVB and self-shielding. However, we find that the radial profiles of the neutral fraction fall smoothly, with median neutral fractions between 30\% and 60\% at $\rhi$ (depending on stellar mass). There is no sharp break in the neutral gas profiles that would set $\rhi$. We have also tested the effect of the UVB explicitly, but the \hi mass-size relation does not change in a simulation that uses the \citet{puchwein_19} UVB prescription.

These considerations hint that the tight \hi mass-size relation arises because the \hi density profiles take on a universal form. The mass-size slope of roughly $0.5$ corresponds to a fixed surface density. Indeed, the surface densities approach more or less constant values at small radii (Fig.~\ref{fig:prof_hi}). Thus, we might speculate that the normalization of the mass-size relation is set by the critical surface density where \hi transitions to \htwo \citep[e.g.][]{bigiel_08}. However, changing this surface density by scaling the UV field in our modelling (see \citetalias{diemer_18_hih2}) does not modify the mass-size relation at all. 

We have also investigated the evolution of the \hi mass-size relation with redshift. Between $z = 0$ and $z = 4$, the relation approximately maintains its slope, but the normalization (the \hi size) decreases with redshift and the scatter increases slightly. In particular, while the TNG100 median relation lies 14\% above the \citet{wang_14_hi} relation at $z = 0$, it agrees to better than 1\% at $z = 1$ and lies below the relation by 9\% and 11\% at $z = 2$ and 4, respectively (see \citet{obreschkow_09a} for similar simulation results). This evolution constitutes a prediction for future observations of the \hi mass-size relation at high redshift.

In summary, the \hi mass-size relation appears to be extremely robust in IllustrisTNG \citep[see also][]{marinacci_17}. The physical reasons for the tight \hi mass-size relation will be further investigated in a forthcoming paper (A. Stevens et al. 2019, in preparation).

\subsection{Anecdotal Evidence from Individual Galaxies}
\label{sec:discussion:individual}

The molecular fraction in the Milky Way is observationally known to be about $\mht / \mh \approx 10^{9} / 10^{10} = 0.1$ \citep[e.g.,][see also \citealt{rachford_02}]{tumlinson_02}. When we select Milky Way analogues with an SFR between $0.5$ and $2 \msun/{\yr}$ and a stellar mass around $5 \times 10^{10} \msun$, the \hiht models predict a median \htwo fraction between $0.3$ and $0.4$. A value of $0.1$ is outside the 68\% scatter of all models, but there are, of course, some galaxies with such relatively low molecular fractions. Moreover, in \citetalias{diemer_18_hih2} we emphasized that our \hiht modelling is systematically uncertain to factors of $2$--$3$.

\citet{zhu_18_malin} recently highlighted a particular galaxy in TNG100 to be an analogue of the massive, low surface brightness gas disc Malin 1 \citep{bothun_87}. \citet{braine_00} found no CO emission in Malin 1, translating into an upper limit of $\mht / \mh < 0.03$, given that Malin 1 contains a large \hi mass of $\mhi = 4 \times 10^{10} \msun$ \citep{pickering_97}. The Malin analogue in TNG100 has a total neutral hydrogen mass of $1.4 \times 10^{11} \msun$, about three times larger than observationally measured. Like in the real Malin galaxy, this gas is distributed in a massive, low surface density disc. Our modelling predicts similarly low molecular fractions between $0.03$ and $0.1$. 


\section{Conclusions}
\label{sec:conclusion}

We have compared the atomic and molecular gas content of galaxies in the Illustris TNG100 and TNG300 simulations to observational data at low redshift. While we largely find good agreement, we uncover a few points of tension. Our main conclusions do not depend on the \hiht model used. They are as follows:
\begin{enumerate}

\item The cosmic abundance of \hi at $z = 0$ is about $8 \times 10^{-4}$ in units of the critical density, roughly twice the abundance measured by 21-cm surveys. The cosmic abundance of \htwo also appears to be slightly higher than observations at $z = 0$, but this disagreement is not conclusive due to selection biases and due to the uncertain CO-to-\htwo conversion factor.

\item The \hi mass function generally agrees well with observations, but overestimates the observed counts around $10^9 \msun$. The \htwo mass function exhibits no significant disagreements with observations.

\item The median \hi and \htwo fractions and their scatter are generally well matched by the simulations. While the \htwo fraction is somewhat low at the highest stellar masses, the uncertainties on the corresponding observations are significant. However, IllustrisTNG produces a sizeable population of satellite galaxies (about 25\% at some stellar masses) that have little or no neutral gas, significantly more than observationally inferred.

\item The relation between \htwo mass and SFR is matched to better than a factor of two in IllustrisTNG, with a constant depletion time of 1 Gyr at $z = 0$. The trend of decreasing depletion time at higher redshift is also reproduced accurately.

\item The observed \hi mass-size relation is matched to 14\% accuracy in IllustrisTNG, and is relatively independent of the \hiht modelling. The radial profiles of \hi surface density also agree well, though a fraction of massive galaxies exhibit excessive central \hi holes. We find indications that the spatial extent of the \htwo distribution is somewhat larger than observed, though with significant scatter.

\item The neutral gas content of simulated galaxies does not show the expected correlations with morphology. As a result, the gas fractions of simulated ETGs are too high, though this statement strongly depends on how morphology is defined.

\item The original Illustris-1 simulation overestimates the neutral gas content of galaxies at $z = 0$ by up to an order of magnitude, highlighting that IllustrisTNG represents an enormous improvement in the galaxy formation model.

\end{enumerate}
It is worth noting that our \hiht results are almost entirely a prediction of the IllustrisTNG simulations. While the model parameters were calibrated to give reasonable gas fractions in cluster halos \citep{pillepich_18_tng}, it is far from obvious that such tuning would automatically result in the correct amount of neutral gas in galaxies over a large range of masses. Similarly, the star formation and feedback models used in the simulations were calibrated to form realistic amounts of stars in halos of a given mass, but it is by no means guaranteed that those stars would form from the right amount of neutral gas. Splitting the gas into \hi and \htwo introduces additional physical demands on the simulation, for example, because \htwo forms in high-density regions. Thus, our \hiht results represent an impressively accurate set of predictions.

However, we have also discovered some areas of tension with observations that can help to inform future generations of cosmological simulations. For example, the sub-grid star formation model of \citet{springel_03} is currently based on a relatively simplistic picture of the temperature and density structure of the ISM, which determines the abundance of cold molecular clouds. Ideally, the ISM model would directly predict the neutral and molecular fractions, rendering post-processing unnecessary. Such a model would be strongly constrained by some of the data we considered in this paper, for instance, by the \htwo-SFR relation and the spatial distribution of \hi and \htwo. Another area where improved observations could constrain simulation physics is the distribution of \hi and \htwo masses at the gas-poor end. As discussed in Section~\ref{sec:discussion:gasfree}, the gas fraction is sensitive to a number physical effects besides star formation and feedback, including the stripping of gas from satellites.

We have left many promising avenues for future work. Most notably, we have focused almost entirely on the local Universe, while there is a rapidly growing body of observations of the gas properties and scaling relations at higher redshift \citep[e.g.,][]{tacconi_13, tacconi_18, genzel_15, freundlich_19}. One particularly interesting question is how molecular gas reservoirs are related to star formation and quenching, and whether their relation changes over redshift. For example, \citet{suess_17} recently reported ALMA detections of two galaxies at $z \approx 0.7$ that host massive reservoirs of molecular gas but exhibit a low SFR. An analysis of similar galaxies in IllustrisTNG could perhaps explain their low star formation efficiency. Finally, we have refrained from a detailed comparison of the spatial distribution of molecular gas, which is increasingly being constrained observationally \citep[e.g.,][]{cormier_16, cormier_18, bolatto_17}.

We emphasize that all simulation data used in this work are publicly available as part of the IllustrisTNG data release \citep{nelson_19_datarelease}. We encourage the community to undertake further analyses and data comparisons.


\section*{Acknowledgements}

We are grateful to Yannick Bah{\'e}, Alberto Bolatto, Martin Bureau, Andreas Burkert, John Forbes, Jonathan Freundlich, Hong Guo, Federico Lelli, Gerg{\"o} Popping, Greg Snyder, Paul Torrey, and Rainer Weinberger for productive discussions. We thank the referee, Romeel Dav{\'e}, for his constructive suggestions. This research made extensive use of the python packages \textsc{NumPy} \citep{code_numpy}, \textsc{SciPy} \citep{code_scipy}, \textsc{Matplotlib} \citep{code_matplotlib}, and \colossus \citep{diemer_18_colossus}. Support for Program number HST-HF2-51406.001-A was provided by NASA through a grant from the Space Telescope Science Institute, which is operated by the Association of Universities for Research in Astronomy, Incorporated, under NASA contract NAS5-26555. ARC acknowledge support from CONACyT graduate fellowship. MV acknowledges support through an MIT RSC award, a Kavli Research Investment Fund, NASA ATP grant NNX17AG29G, and NSF grants AST-1814053 and AST-1814259.


\bibliographystyle{mnras}
\iflocal
\bibliography{../../../../Docs/_LatexInclude/gf.bib}
\else
\bibliography{gf.bib}
\fi


\appendix

\section{Morphological Split}
\label{sec:app:morphology}

\begin{figure}
\centering
\includegraphics[trim = 6mm 9mm 2mm 2mm, clip, scale=0.7]{\figdir/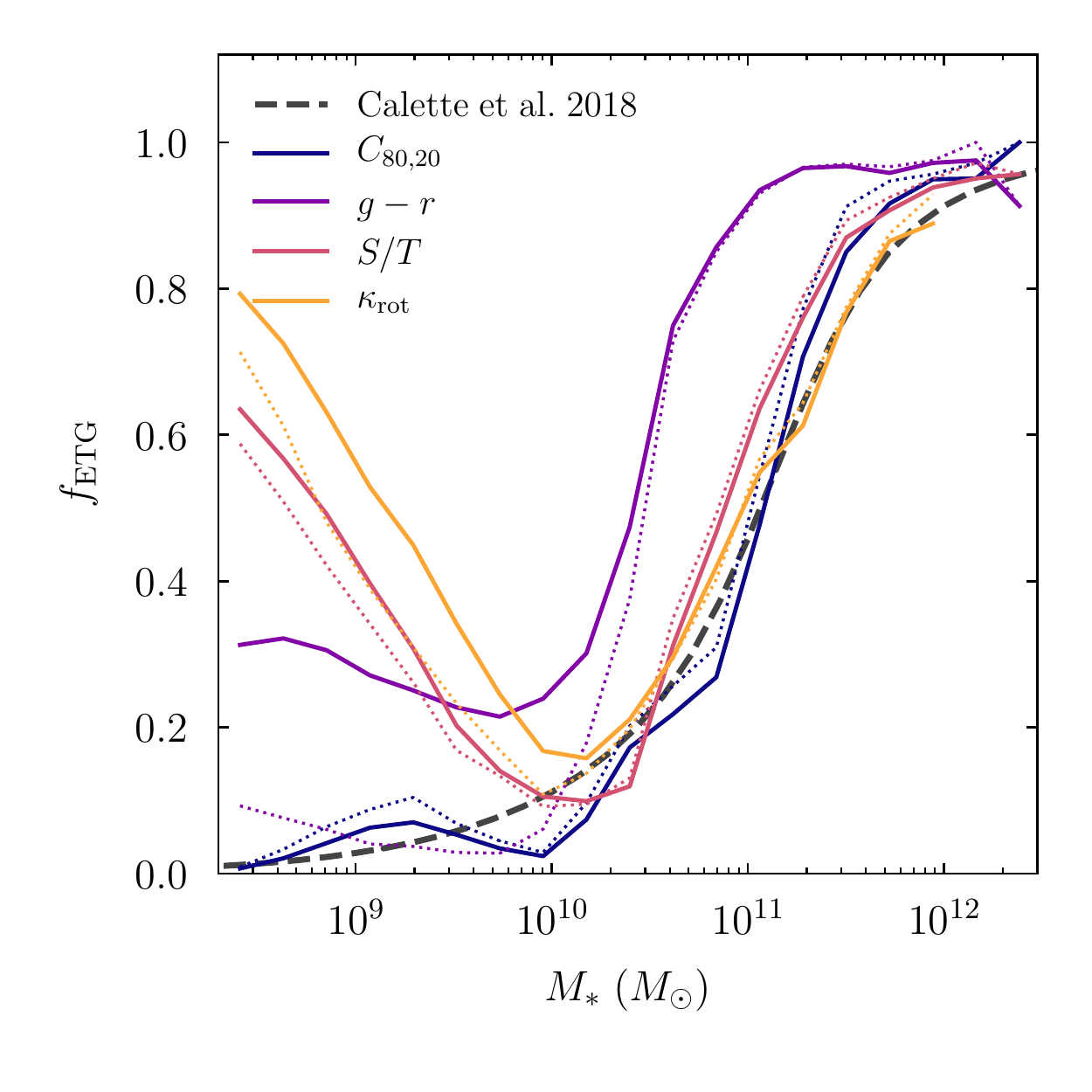}
\caption{The early-type fraction according to a number of pseudo-morphological classifications. The dashed line shows the ETG fraction as a function of stellar mass as parameterized by \citetalias{calette_18}. The solid coloured lines show the fraction in TNG100 and TNG300 when split on stellar concentration (dark blue), g-r colour (purple), spheroid-to-total ratio (red), and the fraction of kinetic energy that is in rotation (yellow). The dotted lines show the fractions if only centrals are considered. The split on stellar concentration does best at matching the \citetalias{calette_18} sample; the similarity of the dotted and solid dark blue lines demonstrates that this split works for both satellites and centrals. We add a scatter of $0.2$ dex to all simulated stellar masses to account for the observational uncertainty (Section~\ref{sec:sim:apertures:star}).}
\label{fig:etgltg}
\end{figure}

\begin{figure*}
\centering
\includegraphics[trim = 1mm 7mm 5mm 0mm, clip, scale=0.55]{\figdir/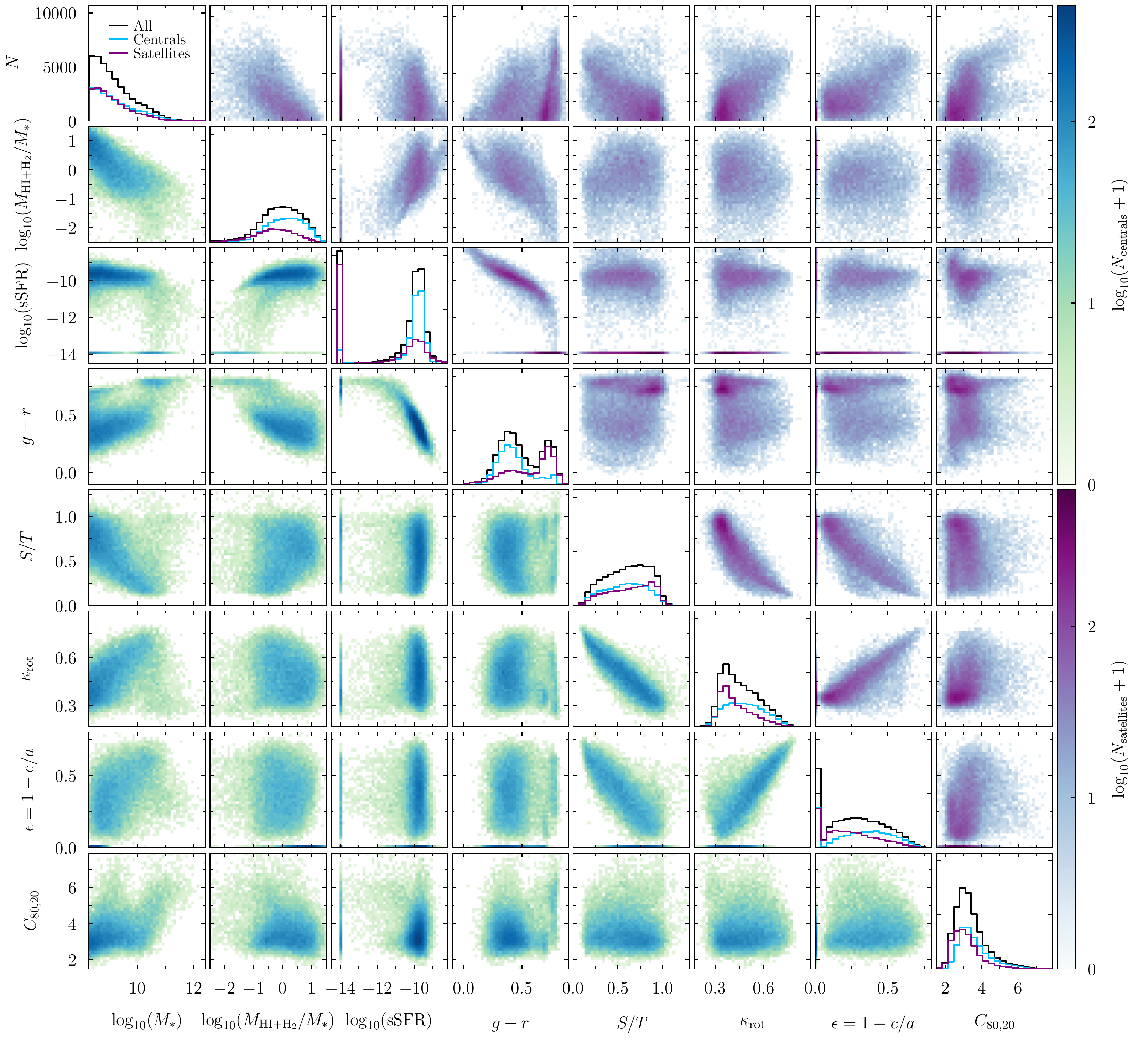}
\caption{Correlations between stellar mass, neutral gas fraction, sSFR, $g$-$r$ colour, spheroid-to-total ratio, fractional energy in rotation, flattening parameter, and stellar mass concentration. The green/blue histograms (lower left half) show the density of central galaxies in TNG100 with $M_* > 2 \times 10^{8} \msun$ in the respective planes of each pair of quantities. The purple histograms (upper right half) show the same for satellites. The histograms along the diagonal show the one-dimensional distributions of each quantity. In IllustrisTNG, the neutral gas content of galaxies correlates much more strongly with colour than with any morphological indicator.}
\label{fig:morphcorrelations}
\end{figure*}

In this appendix, we attempt to categorize IllustrisTNG galaxies according to their morphology in order to investigate the gas content of LTGs and ETGs separately. \citetalias{calette_18} emphasize that the separation into LTGs and ETGs is not entirely uniform within their sample because different methods were applied to the underlying survey data sets, including visual classification and machine learning, which cannot easily be reproduced in simulations \citep[e.g.,][]{huertascompany_11, huertascompany_19}. Nevertheless, we use the ETG fraction as parametrised by \citetalias{calette_18} as a guide. Fig.~\ref{fig:etgltg} shows how splitting on a number of different morphology indicators reproduces this fraction.  Fig.~\ref{fig:morphcorrelations} shows how these indicators correlate with the physical properties of interest, namely, the stellar mass, neutral gas mass, and sSFR.

\setcounter{figure}{0}
\makeatletter 
\renewcommand{\thefigure}{B\@arabic\c@figure}
\makeatother

\begin{figure*}
\centering
\includegraphics[trim =  1mm 9mm 1mm 0mm, clip, scale=0.62]{\figdir/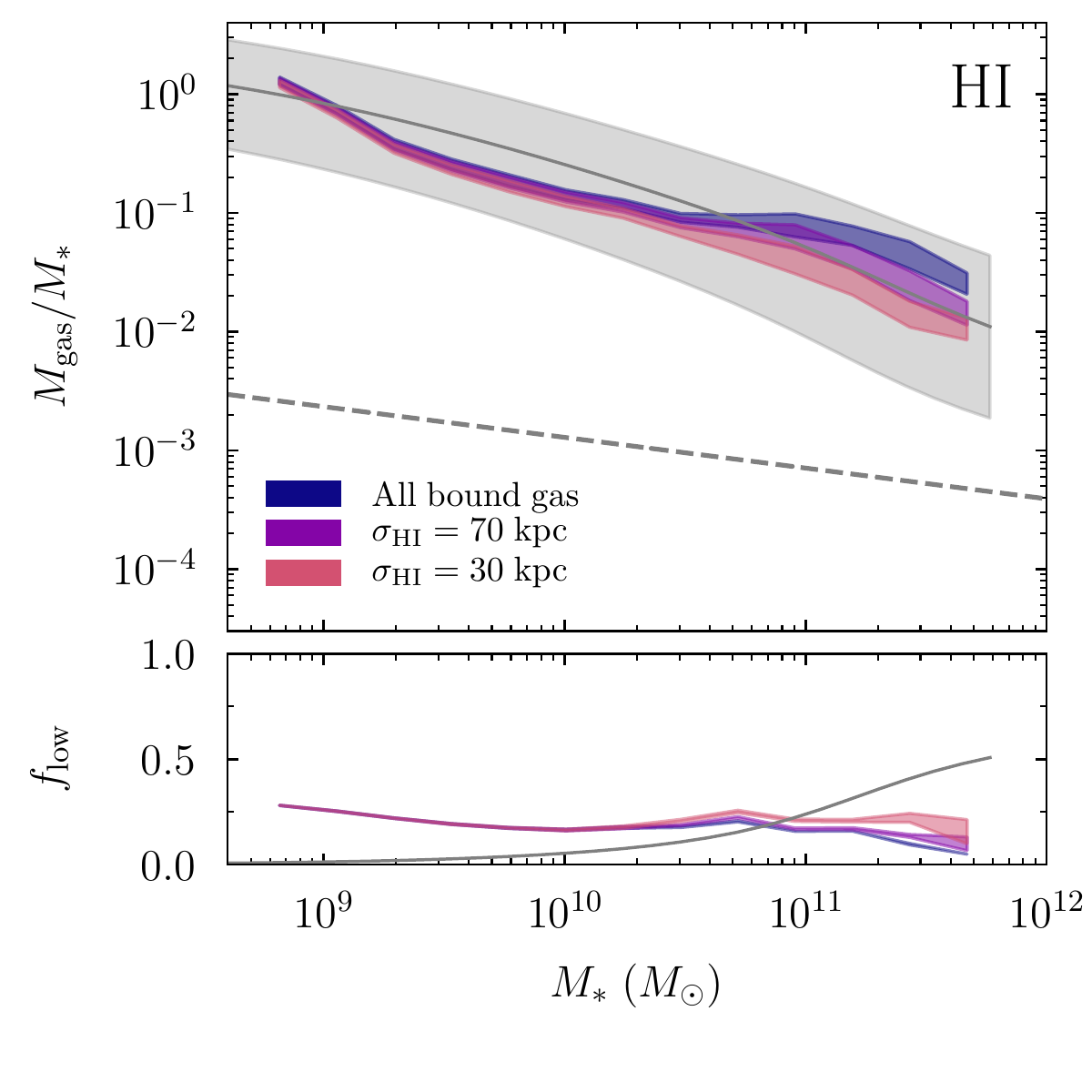}
\includegraphics[trim = 22mm 9mm 1mm 0mm, clip, scale=0.62]{\figdir/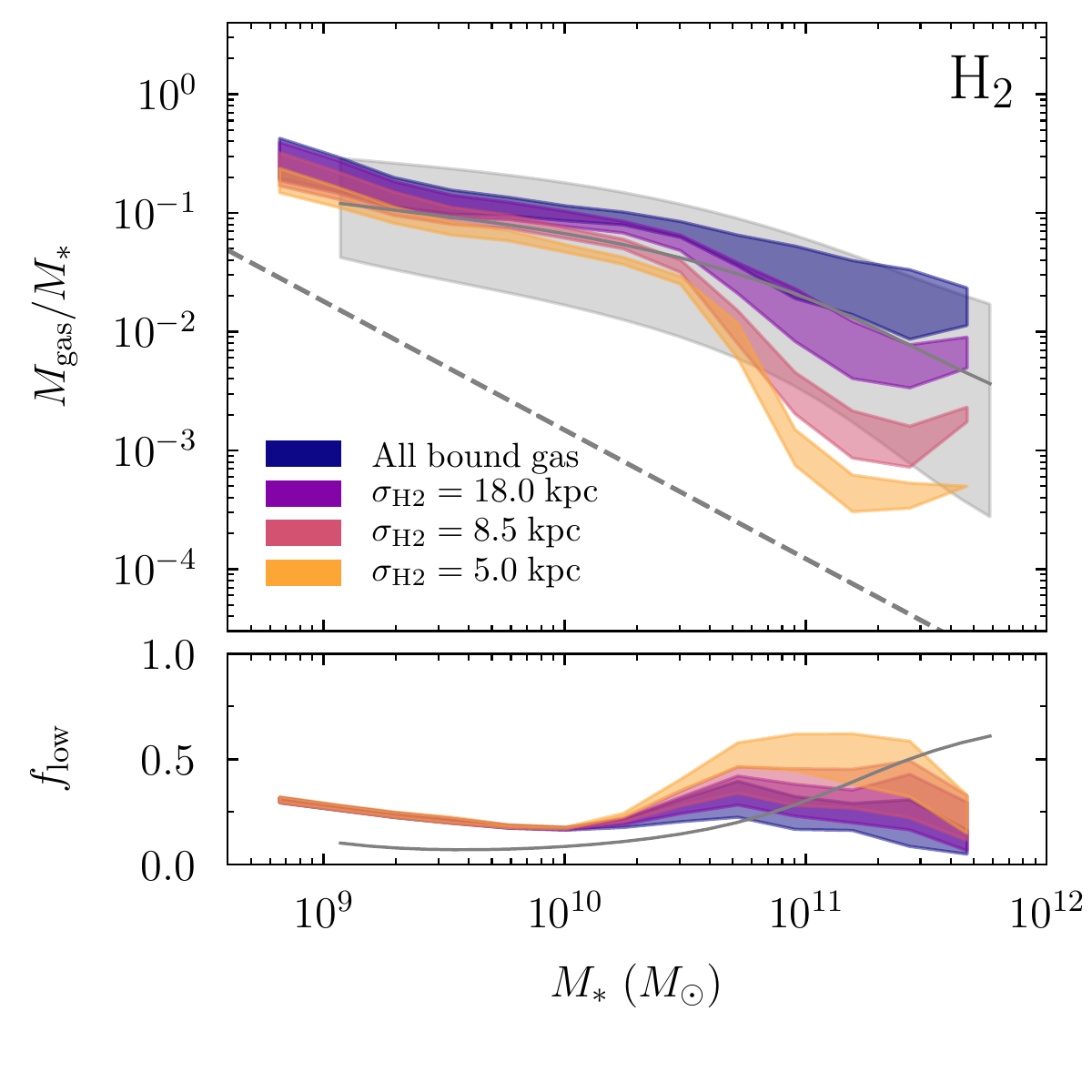}
\caption{Impact of aperture on the mock-observed gas fractions in TNG100. The panels are the same as in Fig.~\ref{fig:fraction}. The shaded regions enclose the different \hiht models, the scatter is not shown to avoid crowding. The dark blue areas show the gas fractions one would infer by including all gas bound to the subhalo as defined by the halo finder. The other colours indicate decreasing apertures, all referring to the standard deviation of a Gaussian beam. At low masses, aperture plays a role only for \htwo which is less spatially extended than \hi. At high masses, our fiducial choices of $\sigma_{\rm HI} = 70$ kpc and $\sigma_{\rm H_2} = 8.5$ kpc lead to a factor of two to three lower \hi masses and up to an order of magnitude reduction in the \htwo mass, highlighting the importance of the aperture choice. The difference between the apertures used for \htwo fractions and the \htwo mass function, $\sigma_{\rm H_2} = 8.5$ kpc and $\sigma_{\rm H_2} = 18$ kpc, is significant.}
\label{fig:gasfrac_aperture}
\end{figure*}

We find that the most suitable indicator of morphology is the concentration of stellar mass, $C_{80,20}$. This concentration can be viewed as a simplified proxy for more complicated profile fitting methods \citep[e.g.,][]{xu_17}. The dark blue lines in Fig.~\ref{fig:etgltg} demonstrate that the criterion $C_{80,20} > 4.9$ reproduces the \citetalias{calette_18} ETG fraction reasonably well. Concentration correlates very weakly with all of the other indicators we tested (Fig.~\ref{fig:morphcorrelations}), indicating that it contains unique information. In the real Universe, however, concentration does correlate significantly with colour \citep{bershady_00}. We have also tested the relationship between $C_{80,20}$ and the bulge statistic, $F(G, M_{20})$ \citep{lotz_04}, which was derived by \citet{rodriguezgomez_18} using an intricate algorithm well matched to those used in observations. $C_{80,20}$ and $F(G, M_{20})$ correlate strongly \citep{tacchella_19}, an indication that $C_{80,20}$ is a suitable criterion that can roughly mimic the morphological classifications performed in the \citetalias{calette_18} sample. When defined in this way, ETGs are dominated by centrals at all stellar masses.

To better understand the connection between gas content and galaxy morphology, we also investigate a number of other pseudo- morphological indicators. First, we consider a cut on the $g$-$r$ colour of IllustrisTNG galaxies, using the dust-corrected colours of \citet{nelson_18_color} wherever available. Fig.~\ref{fig:morphcorrelations} demonstrates that colour correlates strongly with neutral gas fraction and sSFR (the latter more or less by construction). By visual inspection of the colour-mass diagram, we find a split of $g-r = 0.65 + 0.02 [\log_{10}(M_* / M_{\odot}) - 10.28]$ to separate the red and blue galaxy populations. However, this cut leads to a significant excess of ETGs compared to the \citetalias{calette_18} sample, and an increase in ETGs at low masses due to satellites (purple lines in Fig.~\ref{fig:etgltg}). Moreover, colour correlates weakly with structural parameters (Fig.~\ref{fig:morphcorrelations}, see also \citealt{rodriguezgomez_18}), meaning that it is unlikely to be a good proxy for the morphological classifications in observations. On the other hand, when splitting the sample by colour, we obtain a better fit to the ETG and LTG gas fractions of \citetalias{calette_18} than when splitting on concentration. This comparison is clearly somewhat unphysical as the colour split selects the wrong number of ETGs and LTGs (Fig.~\ref{fig:etgltg}), but it highlights that colour correlates more tightly with gas properties than morphology does in IllustrisTNG (Fig.~\ref{fig:morphcorrelations}).

Another type of criterion could be the kinematics of a galaxy's stars, e.g., spheroid-to-total ratio, S/T, or the fraction of kinetic energy in rotation, $\kappa_{\rm rot}$ (as defined in \citealt{rodriguezgomez_17_spin}). We calculate the spheroidal component as twice the mass of all stellar population particles that counter-rotate with respect to the total angular momentum axis of all stars (other definitions lead to very similar outcomes). Fig.~\ref{fig:etgltg} shows the results of a kinematic classification, where ETGs are defined to have $\kappa_{\rm rot} < 0.45$ (yellow) or S/T $> 0.7$ (red). At masses below $M_* \approx 10^{10} \msun$, the ETG fraction increases steeply in both cases, in contrast to the \citetalias{calette_18} sample. Thus, kinematic indicators of morphology cannot, by themselves, be used as a clean indicator of early-type characteristics. 

Finally, we consider indicators of the non-isotropic shape of the stellar distribution, such as the flattening parameter, $\epsilon = 1 - c/a$, and the triaxiality parameter, $T = (a^2 - b^2) / (a^2 - c^2)$. We follow the definitions of \citealt{thob_19}, where $a > b > c$ are the eigenvalues of the inertia tensor of the stellar distribution \citep[computed as in][]{genel_15}. However, there is no straight-forward way to use $\epsilon$ and $T$ for our morphological cut because they are non-monotonic functions of $M_*$ and depend strongly on whether a galaxy is a central or satellite. As expected, $\epsilon$ correlates fairly well with S/T and $\kappa_{\rm rot}$. Splitting the sample according to any of these indicators results in LTG and ETG gas fractions that are more or less similar to a concentration split.

In summary, $C_{80,20}$ appears to provide a reasonable approximation to the ETG fraction of \citetalias{calette_18} and approximates more sophisticated indicators. However, except for colour, the indicators considered do not correlate well with the neutral gas fraction or sSFR of galaxies (Fig.~\ref{fig:morphcorrelations}). This lack of correlation provides the explanation for why the \citetalias{calette_18} gas fractions for LTGs and ETGs are poorly matched compared to the overall sample (Fig.~\ref{fig:gasfrac_c}). This result is in line with the findings of \citet{rodriguezgomez_18}, who showed that the colour-morphology relation does not match Pan-STARRS data, with too many red spirals and blue discs in the simulation. We conclude that the weak correlation between morphology and colour is likely caused by a weak correlation between morphology and gas content.

\section{The Effects of Aperture, Binning, and Blending}
\label{sec:app:apertures}

In Section~\ref{sec:sim:apertures}, we described and justified our choices for the apertures chosen when mock-observing a number of observational surveys.  In this appendix, we investigate the quantitative impact of different aperture choices and show that blending does not have a significant impact on our results. Fig.~\ref{fig:gasfrac_aperture} shows mock-observed \hi and \htwo fractions using a range of apertures. The differences are particularly large for the \htwo fraction at the high-mass end. As discussed in Section~\ref{sec:sim:apertures:h2}, an even smaller aperture of $\sigma_{\rm H_2} = 5$ kpc might be appropriate for the low-redshift galaxies in the xCOLD GASS sample, but such a threshold becomes difficult to interpret because the spatial extent of \htwo gas appears to be broader in IllustrisTNG than in observations (Section~\ref{sec:results:size}). Moreover, reducing the aperture further has a modest impact below stellar masses of $2 \times 10^{10} \msun$. We conclude that our $8.5$ kpc aperture is reasonable but that some systematic uncertainty remains, particularly with respect to the spatial distribution of \htwo. We note that other works have chosen rather different CO apertures. For example, \citet{lagos_15} chose a fixed aperture of $30$ kpc to match their stellar aperture. This choice would increase our mock-observed \htwo masses significantly, bringing them close to the mass of all bound \htwo gas (Fig.~\ref{fig:gasfrac_aperture}).

As we derive aperture masses from projected mass profiles, we have also tested for inaccuracies due to the finite number of radial bins. In particular, we have randomly chosen about $1000$ galaxies and increased the number of bins in their profiles from $50$ to $120$. The stellar masses (aperture $30$ kpc) are changed by less than 1\% for all galaxies. The median \hi masses (aperture $70$ kpc) shift by less than 0.2\% for all \hiht models. In extremely rare cases (less than 2\% of galaxies), the profiles exhibit \hi holes and thus very steep gradients in their cumulative \hi mass profile near the aperture, which can change the mass by about 1\%. The \htwo masses are slightly more affected by binning, with the median \htwo mass changing by less than 5\% for all \hiht models, and added scatter of about 0.15 dex (for an aperture of $8.5$ kpc as used for the gas fraction measurements). When using the $18$ kpc aperture with a hard cut-off for mass functions, the median \htwo mass shifts by between 1\% and 7\% depending on the \hiht model. None of these shifts are relevant to the conclusions of this paper.

Finally, we check the importance of blending. \citet{stevens_19_hi} carefully considered the redshift distribution of their observed \hi samples and modelled the effect of blending, i.e., the confusion of multiple \hi sources in a single beam. We have compared our modelling to theirs for a representative sample of galaxies with $M_* > 10^9 \msun$. The following results refer to the volumetric \modelgk model, but the results do not strongly depend on the \hiht model. We find that our \hi masses (aperture $70$ kpc) agree with the \citet{stevens_19_hi} sample with a median ratio of $0.97$ and a 68\% interval between $0.72$ and $1.06$, for all galaxies that host significant \hi in both samples. There is, however, a population of 4\% of the galaxies that exhibit significant \hi content ($\mhi > 10^8 \msun$) in the \citet{stevens_19_hi} sample but no \hi ($\mhi < 10^6 \msun$) in our sample --- some of the gas-free objects discussed in Section~\ref{sec:discussion:gasfree}, which experience \hi contributions from gas in other galaxies. Generally, blending most strongly affects satellites, but our sample is dominated by centrals across the entire mass range. Thus, the statistical shift in our \hi masses compared to \citet{stevens_19_hi} is acceptable.


\bsp
\label{lastpage}
\end{document}